\documentclass[fleqn,usenatbib]{mnras}

\usepackage[T1]{fontenc}

\DeclareRobustCommand{\VAN}[3]{#2}
\let\VANthebibliography\thebibliography
\def\thebibliography{\DeclareRobustCommand{\VAN}[3]{##3}\VANthebibliography}

\usepackage{graphicx}	
\usepackage{amsmath}	
\usepackage{amsfonts}
\usepackage{amssymb}	
\usepackage{color}
\usepackage[normalem]{ulem} 
\usepackage{newtxtext,newtxmath}

\def\bm#1{\mbox{\boldmath$#1$}}

\title[Evolutionary stage of Betelgeuse]{The evolutionary stage of Betelgeuse inferred from its pulsation periods}

\author[H. Saio et al.]{
Hideyuki Saio,$^{1}$\thanks{E-mail: saio@astr.tohoku.ac.jp (HS)}
Devesh Nandal,$^{2}$
Georges Meynet$^{2}$
and Sylvia Ekst\"om$^{2}$
\\
$^{1}$Astronomical Institute, Graduate School of Science, Tohoku University, Aramaki, Aoba-ku, Sendai, Miyagi, 980-8578, Japan\\
$^{2}$D\'epartement d\'Astronomie, Universit\'e de Gen\'eve, Chemin Pegasi 51, CH-1290 Versoix, Switzerland
}

\date{Accepted XXX. Received YYY; in original form ZZZ}

\pubyear{2023}

\begin{document}
\label{firstpage}
\pagerange{\pageref{firstpage}--\pageref{lastpage}}
\maketitle

\begin{abstract}
Betelgeuse is a well known bright red supergiant that shows semi-regular 
variations with four approximate periods of 2200, 420, 230, and 185 days. 
While the longest period was customarily regarded as 
LSP (long secondary period) of unknown origin, 
we identify the $\sim$2200-d period as the radial fundamental mode, and 
the three shorter periods as the radial first, second, and third overtones. 
From a linear nonadiabatic pulsation analysis including the pulsation/convection coupling,
we have found that these radial pulsation modes are all excited in the
envelope of a model in a late stage of the core-carbon burning.
Models with similar pulsation property
have masses  of $11\sim12\,M_\odot$ ($19\,M_\odot$ at ZAMS) with 
luminosities ($\log L/L_\odot =5.27\sim5.28$) and effective temperatures
($\log T_{\rm eff}\approx 3.53$) that are  
consistent with the range of the observational determinations.
We also find that a synthetic light curve obtained by adding 
the fundamental and the first-overtone mode is comparable
with the light curve of Betelgeuse up to the Great Dimming. 
We conclude that Betelgeuse is likely in the late stage of core carbon burning, 
and a good candidate for the next Galactic Type II supernova.  
\end{abstract}




\section{Introduction}

Betelgeuse ($\alpha$ Orionis; HD\,39801) is a nearby luminous red supergiant 
(RSG; M1-2 Iab-a) located approximately 200 parsec from
the Sun, has a fascinating past. 
For example, from investigating of pre-telescopic records
\citet{Neuhauser2022} found that its surface temperature was significantly 
higher two millennia ago. 
Betelgeuse shows semi-periodic light variations, which have been observed 
for more than a century (AAVSO\footnote{The AAVSO International Database, https://www.aavso.org}).
These and other very interesting phenomena that occurred in and around Betelgeuse are nicely reviewed in \citet{Wheeler2023}.

The brightness of Betelgeuse decreased abnormally (called `Great Dimming') 
from  December 2019 to February 2020 (by $\sim\!1.2$\,mag) 
around a minimum epoch of $\sim400$\,d variation \citep{Guinan2019}.
Stimulated by the mysterious dimming, many intensive observations 
in various wavelengths, and detailed analyses have been carried out. 
These investigations suggest that effective temperature decreased by $\sim100$\,K
\citep[e.g.,][]{Dharmawardena2020,Harper2020,Taniguchi2022,Wasatonic2022,Mittag2023}
during the Great Dimming, and a substantial mass ejection possibly occurred
\citep[e.g.,][]{Montarges2021,Kravchenko2021,Taniguchi2022,Dupree2022,Jadlovsky2023}. 

Given Betelgeuse's high luminosity and its complex variations, 
which could potentially indicate an impending supernova event, the evolution
of this star has been the subject of numerous investigations
 \citep[e.g.,][]{Meynet2013,Dolan2016,Wheeler2017,Nance2018, Luo2022}.
Its high rotation rate $\varv_{\rm eq} = 6.5^{+4.1}_{-0.8}$km\,s$^{-1}$
\citep{Kervella2018} as a red supergiant has stimulated discussions on the evolution
involving merger with a past companion \citep{Sullivan2020} 
as a source of angular momentum. 

Pulsation periods of the semi-regular variations of Betelgeuse are very useful
to constrain its present fundamental parameters and the evolution stage.
Periods of $2200$\,d and $420$\,d are obvious and most frequently mentioned.
So far, the $420$-d period was customarily considered to be the radial 
fundamental mode \citep[e.g.][]{Joyce2020}, and the $2200$-d period to be a LSP (Long Secondary Period)
caused by something other than radial pulsations.
However, we will show, in this paper, that in a luminous supergiant like Betelgeuse
long-period pulsations are very nonadiabatic so that $2200$-d period is
suitable for the radial fundamental mode,
and the $420$-d period should be identified as the first overtone mode. We can confirm this mode identifications by extending the period-luminosity (PL) relations of less luminous red-giant (RG) to the RSG range in Fig.7 of \citet{Kiss2006}
\citep[or Fig.6 of][]{Chatys2019}. We find that the $420$-d period at $M_{\rm K}\approx -10$\,mag ($M_{\rm K}=$ absolute magnitude in K band) lies on the extension of sequence B which is identified as the first overtone radial pulsation \citep[e.g.,][]{Trabucchi2017}. This clearly indicates the $420$-d period to be the first overtone mode of Betelgeuse.  
Since the pulsation period of a given mode is shorter for a smaller radius, 
the size of Betelgeuse would be significantly underestimated if the $420$-d period 
were fitted with radial fundamental mode (rather than first overtone).

\section{Observational constraints}

\subsection{Global parameters}

\citet{Lobel2000} obtained the surface gravity $\log g = -0.5$ and 
effective temperature $T_{\rm eff} = 3500\pm100$\,K by fitting of 
synthetic profiles to unblended metal absorption lines in the near-IR spectra.
Furthermore, from CO equivalent widths for Betelgeuse \citet{Carr2000} obtained 
$T_{\rm eff}=3540\pm260$\,K.
More recently, \citet{Levesque2020} obtained $T_{\rm eff} = 3600 \pm 25$\,K from
optical spectrophotometry.
Based on these spectroscopic $T_{\rm eff}$ determinations for Betelgeuse, we adopt
$T_{\rm eff} = 3500\pm200$\,K for a constraint for our models, and note that 
the $T_{\rm eff}$ range is the same as that adopted by \citet{Dolan2016}. 

As emphasized in \citet{Josselin2000}, K-magnitude is more useful in deriving 
bolometric luminosity of Red supergiants (RSG) rather than V-magnitude,
because the K-bolometric correction is insensitive to  
the effective temperature and surface gravity. 
In addition, K-magnitude is less affected by interstellar extinction or pulsation.
From the 2MASS All-Sky Catalog of Point Sources \citep{Cutri2003},
K-magnitude of Betelgeuse is $-4.378\pm0.186$ mag.
Adopting the distance of $222^{+48}_{-34}$\,pc from the new combined 
radio+Hipparcos astrometric solution obtained by \citet{Harper2017}, 
and K-bolometric correction  $2.92\pm0.16$ \citep{Levesque2005}
for $T_{\rm eff}=3500\pm200$\,K,
we obtain $\log L/L_\odot = 5.18 \pm 0.11$ for Betelgeuse \citep[which is similar
to $5.10\pm0.22$ adopted by][]{Dolan2016}.

\subsection{Observed pulsation periods}
\label{sec:puls} 

\begin{table*}
	\centering
	\caption{Recent period determinations for light and radial-velocity curves of Betelgeuse and the adopted periods in this paper }
	\label{tab:periods}
	\begin{tabular}{rllllll} 
      \hline 
      Authors & \multicolumn{5}{c}{Periods (d)} & Data\\
      \hline
		\citet{Jadlovsky2023}: & $2190 \pm 270$ & $417 \pm 17$ & $365 \pm 75$ & $230 \pm 29$ & $185 \pm 4$ &AAVSO$^{1)}$,SMEI$^{2)}$\\
	\citet{Ogane2022}: & $2160$ & $405$ && $202^{\rm a)}$ & 
	& UBVRI photometry\\
	\citet{Wasatonic2022}: & $2209 \pm 183$ & $439 \pm 5$ & & & & V \& NIR photometry\\
     \citet{Granzer2022}: & $2169\pm 6.3$ & $394.5$ & & $216.0$ & & STELLA$^{3)}$\\
	 \citet{Joyce2020}: & $2365 \pm 10$ & $416\pm24$ &&& $185.5\pm0.1$ & AAVSO,SMEI\\
	\citet{Kiss2006}: & $2050\pm460$ & $388\pm 30$ &  &  & & AAVSO\\
      \hline
       & $P_1$ & $P_2$ & & $P_3$ & $P_4$ \\
      Adopted in this work: & $2190 \pm 270$ & $417 \pm 24$ & & $230 \pm 29$ & $185 \pm 4$ \\
		\hline
	\end{tabular}\\
 $^{1)}$ AAVSO= American Association Vatiable Star Observers, \quad
 $^{2)}$ SMEI= Solar Mass Ejection Imager \quad
 $^{3)}$ STELLA= \'echelle spectrograph \\
 $^{\rm a)}$ If this periodicity is subtracted from the V-band data 
 as the harmonic of $405$\,d,
additional periods of $237.7$\,d and $185.8$\,d are obtained. 
\end{table*}

The semi-regular light variations of Betelgeuse indicate several periodicities 
ranging from $\sim2200$\,d to $185$\,d to be simultaneously excited.
Indeed, every period analysis for Betelgeuse light curve yielded more 
than two periods
\citep{Jadlovsky2023,Ogane2022,Wasatonic2022,Granzer2022,Joyce2020,
Kiss2006,Stothers1971}.
Table\,\ref{tab:periods} lists most recent period determinations by various authors. 
Among five groups of periods, we have selected the four periods $P_1$,$P_2$,$P_3$ 
and $P_4$ as listed in the last line of Table\,\ref{tab:periods}.
(The 365-day period of \citet{Jadlovsky2023} is considered to be one year alias.)
Regarding these four periods being caused by radial pulsations of Betelgeuse,
we have searched for evolution models which simultaneously excite pulsations 
of the four periods $P_1, \ldots P_4$ in the range of the
error box for Betelgeuse in the HR diagram.

\section{Models}

Evolution models of massive stars from ZAMS (Zero-age main-sequence) 
up to the end of core-carbon burning (to the end of silicon burning for a 
selected case) have been computed using the recent version of 
the Geneva stellar evolution code \citep[GENEC:][]{Ekstroem2012,Yusof2022}.  
We have assumed the same initial chemical composition of \citet{Yusof2022}; 
i.e., initial mass fractions of hydrogen and heavy elements 
are set as $X=0.706,Z=0.020$.
The mixing length to pressure scale hight is set $1.6$ for the envelope convection.
Convective core boundaries are determined using the Schwarzschild criterion 
with a step overshooting of $0.1\,H_p$ with $H_p$ being the pressure scale height 
at the boundary.

We adopt initial rotation velocities, $\varv_{\rm i}$ of $0.1, 0.2$, or 
$0.4\,\varv_{\rm crit}$ with $\varv_{\rm crit}$ being the critical rotation 
 velocity at ZAMS.
While the resulted rotation velocities in the envelopes of the red supergiants 
are too small to impact the radial pulsations, rotational mixing 
increases effective core mass and hence luminosity \citep{Ekstroem2012,Yusof2022}. 
Furthermore, CNO abundances at the surface are affected by 
the assumed initial rotation rates.  

We have obtained periods and stability (i.e, excitation/damping) of radial 
pulsations of various models using the same nonadiabatic linear radial pulsation 
code employed in \citet{Anderson2016} for Cepheid pulsations.
The pulsation code is based on that described in \citet{Saio1983} but revised
to include the effect of the coupling between pulsation and time-dependent convection. 
As summarized in Sec.2.2 of \citet{Anderson2016}, the convection effects 
are included based on the works by \citet{Unno1967} and \citet{Liege2005}, 
which successfully predicts the red-edge of the cepheid instability strip.
Details of the pulsation code are discussed in Appendix A. 

\section{Results}
\subsection{Periods of excited radial pulsations}
\begin{figure}
	\includegraphics[width=\columnwidth]{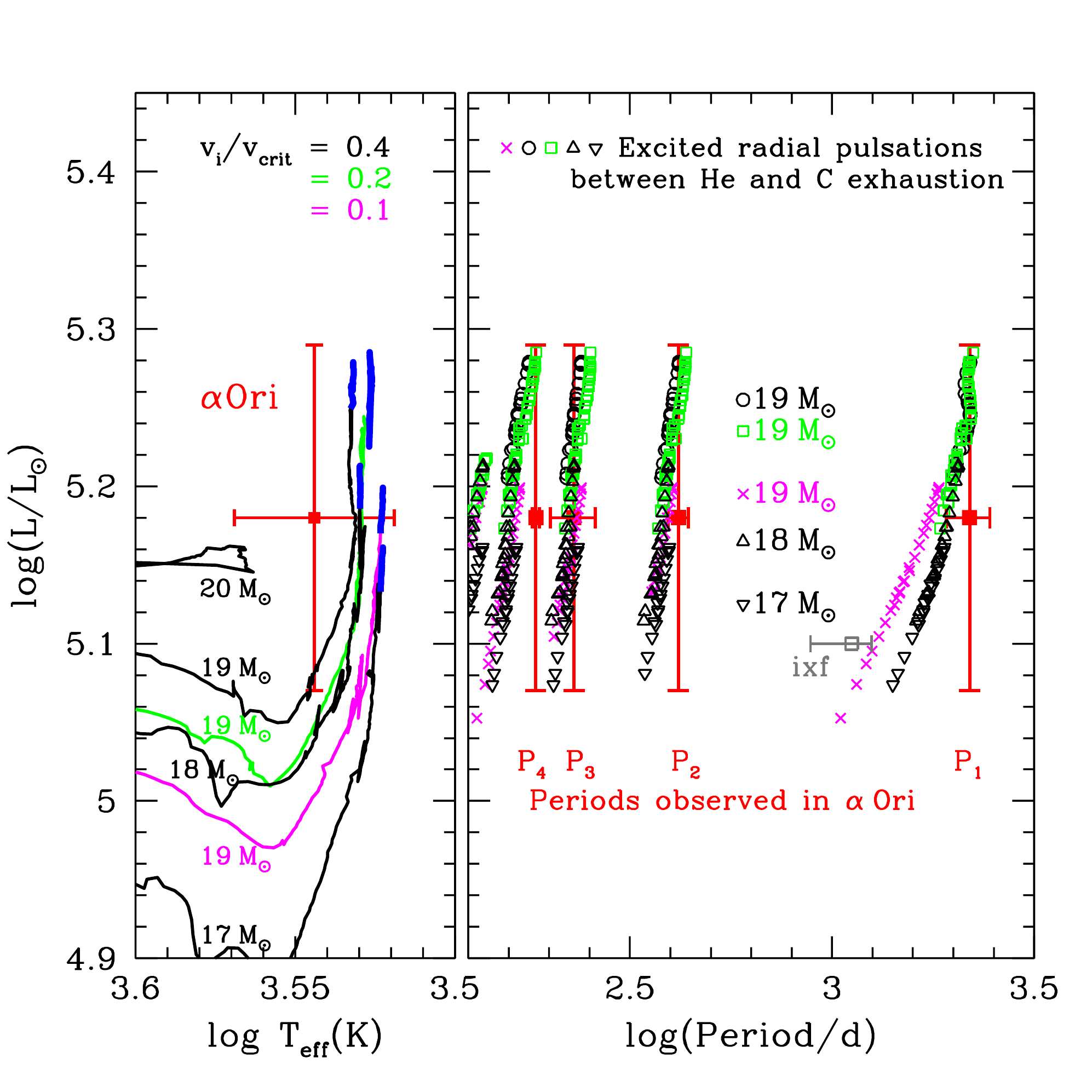} 
    \caption{{\bf Left panel:} The location
     of Betelgeuse ($\alpha$ Ori) and various
     evolutionary tracks on the HR diagram.
     The blue-coloured sections represent the phase of core-carbon burning.
      The number along each track indicates the initial mass in solar units. 
      The $20\,M_\odot$-track is taken from \citet{Yusof2022}.
    {\bf Right panel:} Period-luminosity diagram showing observed four periods of Betelgeuse with error bars (red lines), and various symbols for excited radial-pulsation periods along the evolutionary tracks from the He exhaustion to the Carbon exhaustion at the stellar center.
    For comparison, the pulsation period of the RSG progenitor of SN2023ixf (in M101) obtained by \citet{Jencson2023} is shown by the gray horizontal bar labelled as "ixf". 
	}
  \label{fig:te_lum_peri}
\end{figure}

\begin{table*}
	\begin{center}
	\caption{Model examples which excite pulsations consistent with periods of Betelgeuse }
	\label{tab:models}
	\begin{tabular}{ccccccccccccc} 
      \hline 
  Model &   $M_{\rm i}^{\rm a)}$ & $\varv_{\rm i}/\varv_{\rm crit}$ & $P_1$[d] & $P_2[d]$ & $P_3[d]$ & $P_4[d]$ & $M^{\rm a)}$ & $\log L/L_\odot$ & $\log T_{\rm eff}[K]$ & $\log R/R_\odot$& $X_{\rm c}({\rm C})^{\rm b)}$ \\ 
    $\alpha$Ori   &  &  & $2190 \pm 270$ & $417 \pm 24$ & $230 \pm 29$ & $185 \pm 4$ &  &
         $5.18\pm0.11$ & $3.544\pm 0.025$ \\ 
      \hline
   A &   19 & 0.4 & 2199 & 418 & 240 & 178 & 11.23 & 5.279 & 3.532 & 3.100& 0.0067 \\ 
      \hline
   B &   19 & 0.2 & 2186 & 434 & 252 & 184 & 11.73 & 5.276 & 3.526 & 3.109 
   & 0.0048  \\ 
   C &   19 & 0.2 & 2181 & 434 & 252 & 184 & 11.73 & 5.275 & 3.526 & 3.109 
   & 0.0503  \\ 
  D &  19 & 0.2 & 2168 & 428 & 249 & 181 & 11.73 & 5.265 & 3.526 & 3.103  
  & 0.1712 \\ 
     \hline
     	\end{tabular}\\
 \end{center}
 \leftline{$^{\rm a)}$ Initial mass $M_{\rm i}$ and present mass $M$ in the solar unit.}
 \leftline{$^{\rm b)}$ Mass fraction of carbon at the center.} 	
\end{table*}

\begin{figure}
\includegraphics[width=\columnwidth]{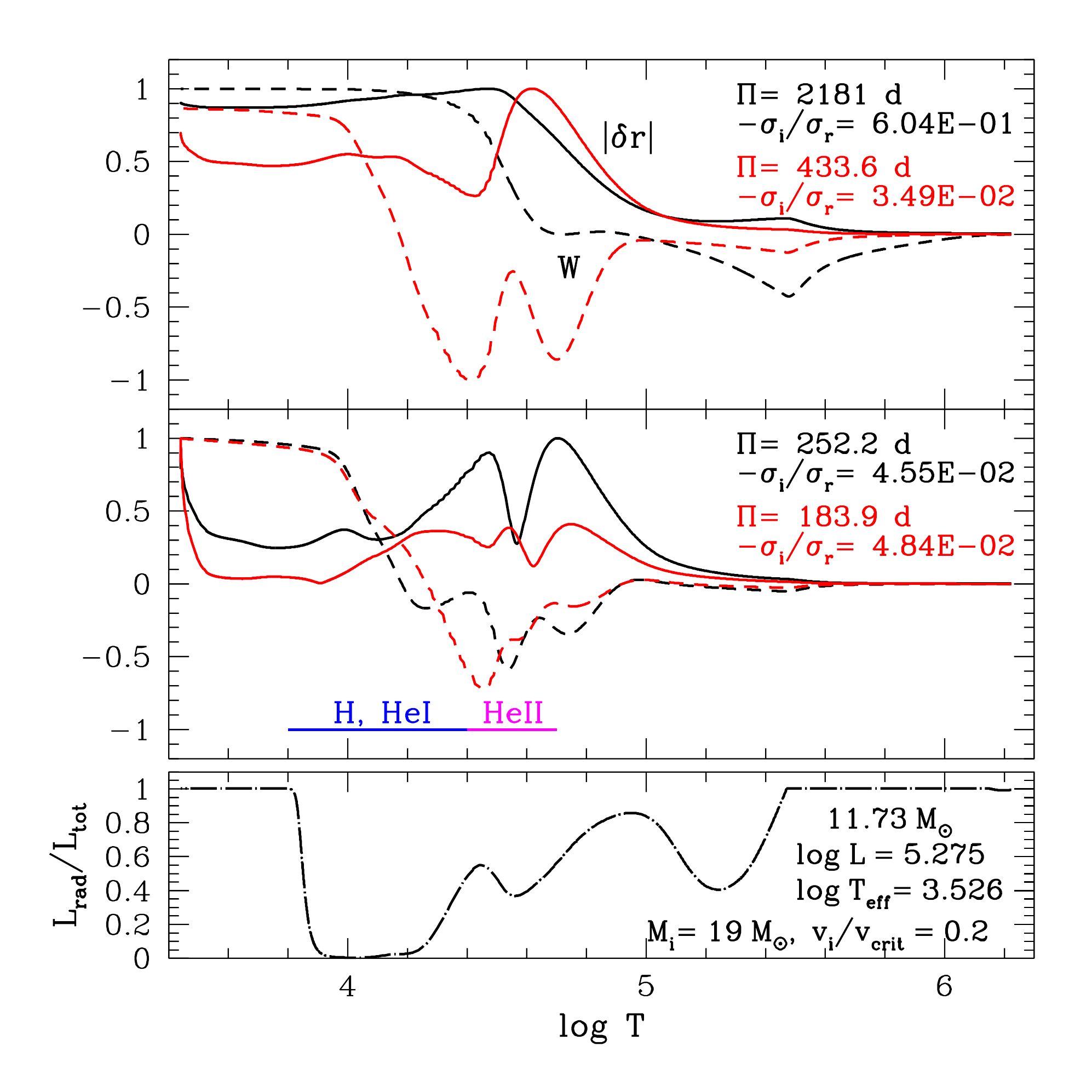} 
\caption{Pulsation properties of excited modes (upper panels)
  and radiative luminosity (bottom panel) versus 
 logarithmic (base 10) temperature for model C (Table\,\ref{tab:models}).
The top and middle panels show the absolute value of radial displacements
 $|\delta r|$ normalized to their maximum value (solid lines) 
 and (cumulative) work curves $W$ (dashed lines).
The ratio of imaginary to real parts of eigenfrequency $\sigma_i/\sigma_r$
gives growth (if $\sigma_{\rm i} < 0$) or damping ($\sigma_{\rm i}>0$) rates. 
The horizontal line in the middle panel indicates 
H,HeI (blue) and HeII (magenta) ionization zones.
Red color is used to distinguish different modes plotted in one panel. 
The bottom panel shows the variation of the radiative to total 
luminosity ratio $L_{\rm rad}/L$ in the envelope. 
}
\label{fig:xiwork}
\end{figure}

We have applied our non-adiabatic radial pulsation code to selected evolution
models which enter the error box on the HR diagram (Figure\,\ref{fig:te_lum_peri}).
We have found that the four radial pulsation modes having periods similar to those 
observed in Betelgeuse  (right panel of Fig.\,\ref{fig:te_lum_peri}) are excited in models located in the cooler and luminous 
part of the error box in the HR diagram as shown in Figure\,\ref{fig:te_lum_peri}.
Table\,\ref{tab:models} lists good models in which all 
the excited radial pulsation periods agree with the observed four periods.
These models are in the carbon-burning (or ending) phase having luminosities
in the range of $5.26 < \log L/L_\odot < 5.28$. 
The mass at ZAMS was $19\,M_\odot$ and has been reduced to  
$\sim11 - 12\,M_\odot$ at the phase of Betelgeuse. 

Note that we need a high luminosity in the upper range of error bar, 
because the predicted period of the third overtone mode deviates from $P_4$ quickly 
as the luminosity decreases, while $P_1$ varies slowly  as seen in the right panel of Figure\,\ref{fig:te_lum_peri}. This property of $P_4$ sets the lowest luminosity acceptable among models with $\varv_{\rm i}=0.2\varv_{\rm crit}$ at model D (Table\,\ref{tab:models}) which has $\log L/L_\odot=5.265$ and $P_4=181$\,days at the shortest limit of the observed range $185\pm4$\,days. The carbon abundance in the core of model D is 0.171, which will be exhausted in 260 years. At every stage of evolution from model D to B (carbon exhaustion), periods of excited pulsation modes agree with those of Betelgeuse. Model C is chosen as a typical model between B and D.

For the models of $\varv_{\rm i}=0.4\varv_{\rm crit}$ we have listed only Model A in Table\,\ref{tab:models} although the $P_4=178$\,days is slightly shorter than the observational limit (i.e., a slightly larger luminosity is needed to perfectly agree with $P_4$). The central carbon in model A is almost exhausted, and the luminosity and radius do not change any more to the carbon exhaustion phase.

We note that in the lower luminosity range
 $\log L/L_\odot < 2.52$, an 
additional shorter period mode (fourth overtone) is excited
(right panel of Fig.\,\ref{fig:te_lum_peri}), but it is not observed,
 which additionally supports a high luminosity of Betelgeuse.

The spatial amplitude variation of each excited mode in model C as an example 
is shown in Figure\,\ref{fig:xiwork} (top and middle panels) 
by solid lines as a function of temperature in the envelope. 
In calculating linear pulsation modes we solve a set of equations 
with an eigenvalue $\sigma$, expressing the temporal variation of 
radial displacement as $\delta r\!\exp(i\sigma t)$, in which
$\sigma$ and $\delta r$ (sometimes called the eigenfunction) 
are a complex number and a complex spatial function, respectively.
The actual pulsation is expressed as the real part 
$[\delta r\!\exp(i\sigma t)]_{\rm r}$
so that the imaginary part of $\delta r$ expresses the deviation from the 
standing wave character in the envelope (i.e., zero points of displacement
shift during a cycle of pulsation), and $\sigma_{\rm i}$ 
(imaginary part of $\sigma$) gives the damping/growth rate 
of the pulsation mode. 
In contrast to the nearly adiabatic pulsations of the classical Cepheids,
pulsations of supergiants are very non-adiabatic so that
the imaginary parts of $\delta r$ and $\sigma$ can be comparable to 
the real parts.

The top panel of Figure\,\ref{fig:xiwork} shows that $\delta r$ 
of the longest period $P_1$ ($2181$\,d) mode is nearly flat in the
outermost layers decreasing gently toward the center, which is exactly the
property of radial fundamental mode. 
For this reason we identify $P_1$ as the fundamental mode (rather than LSP).
The displacement of $P_2$ ($434$\,d) mode has a dip at $\log T\approx 4.4$
corresponding to a node in the adiabatic pulsation, which indicates
$P_2$ to be the first overtone mode. 
No clear nodes appear in strongly nonadiabatic pulsations, 
because zero points of real and imaginary parts of 
$\delta r$ are separated.
In the middle panel, 
$\delta r$ (solid lines) $P_3$ and $P_4$ change rapidly with a few dips. 
We identify $P_3$ and $P_4$ as second and third overtones, respectively.
We note that all pulsations are confined in the envelope of $\log T < 6$.

Note that $|\delta r|$'s for short period modes (for $P_3$ and $P_4$) in the 
middle panel of Figure\,\ref{fig:xiwork} steeply change near the surface.
This indicates that the pulsation energy of these modes leaks 
at the surface, for which we apply running-wave outer boundary condition
in solving the eigenvalue problem.
In spite of the leakage of energy, we find that these pulsations 
still grow (i.e., $\sigma_{\rm i} < 0$) because the driving in the inner part
exceeds the energy loss at the surface.

\subsection{Work integrals}
A dashed line in the top or middle panel of Figure\,\ref{fig:xiwork}  
presents the cumulative work $W$ of each mode. 
The sign of the gradient $dW/dr$ indicates whether the layer locally 
drives ($dW/dr>0$) or damps ($dW/dr<0$) the pulsation.  
Furthermore, whether $W$ at the surface is positive or negative means
the pulsation to be excited or damped in the model, respectively. 
The cumulative work $W(r)$ is calculated as \citep[e.g.,][]{SaioCox1980} 
\begin{equation}
W(r) \propto \int_0^r r^2P\,{\rm Im}\left({\delta\rho\over\rho}{\delta P^*\over P}\right) dr,
\label{eq:work}
\end{equation}
where $P$ and $\rho$ are pressure and density, respectively, while $\delta$ means
 the  pulsational perturbation of the next quantity, and Im$(\ldots)$ and $(^*)$ 
 mean the imaginary part and the complex conjugate, respectively. 
Figure\,\ref{fig:xiwork} indicates that the fundamental mode $P_1$ 
of Betelgeuse is mainly driven in the HeII ionization zone at $\log T\sim4.5-4.6$,
while for other shorter period modes the driving in the H/HeI ionization zone  
($4.0\la\log T\la4.4$) is most important.  

The bottom panel of Figure\,\ref{fig:xiwork} shows the ratio of radiative to 
total luminosity, $L_{\rm rad}/L_{\rm tot}$, which is considerably smaller 
than unity in large ranges of the envelope indicating a significant fraction of
the total energy flux to be carried by convection.
The radiative/convective flux ratio affects the
period as well as the driving/damping of the pulsation of a star
with high luminosity to mass ratio, $L/M > 10^4$ (in solar units). 
Because the thermal time is comparable or shorter than
the dynamical time in the envelope of such a luminous star,
the pulsational variation of energy flux plays an important role to
determine the pulsation period not only to excite or damp the pulsation. 
In fact, if we obtain pulsation periods of model C (Table,\ref{tab:models}) 
neglecting convective flux perturbations (which is sometimes called
frozen convection approximation) we obtain periods of 1771, 409, 253, and 192 days.
Comparing these periods with periods from the full calculations in model C,
2181, 434, 252, and 184 days, we find including convection flux perturbation to be 
important only for longer period pulsations in such luminous stars.
This indicates that if we fitted the observed $2200$-d period with the fundamental period obtained without convection effects, we would have a $\sim17\%$ larger radius, because the pulsation period is approximately proportional to $R^{1.5}$.
In addition, the fact that the convection effect is serious only in the fundamental mode suggests that even with a pulsation code without convection effects, we would have comparable asteroseismic models by using only overtone modes (i.e., avoiding the fundamental mode). 

\begin{figure}
\includegraphics[width=\columnwidth]{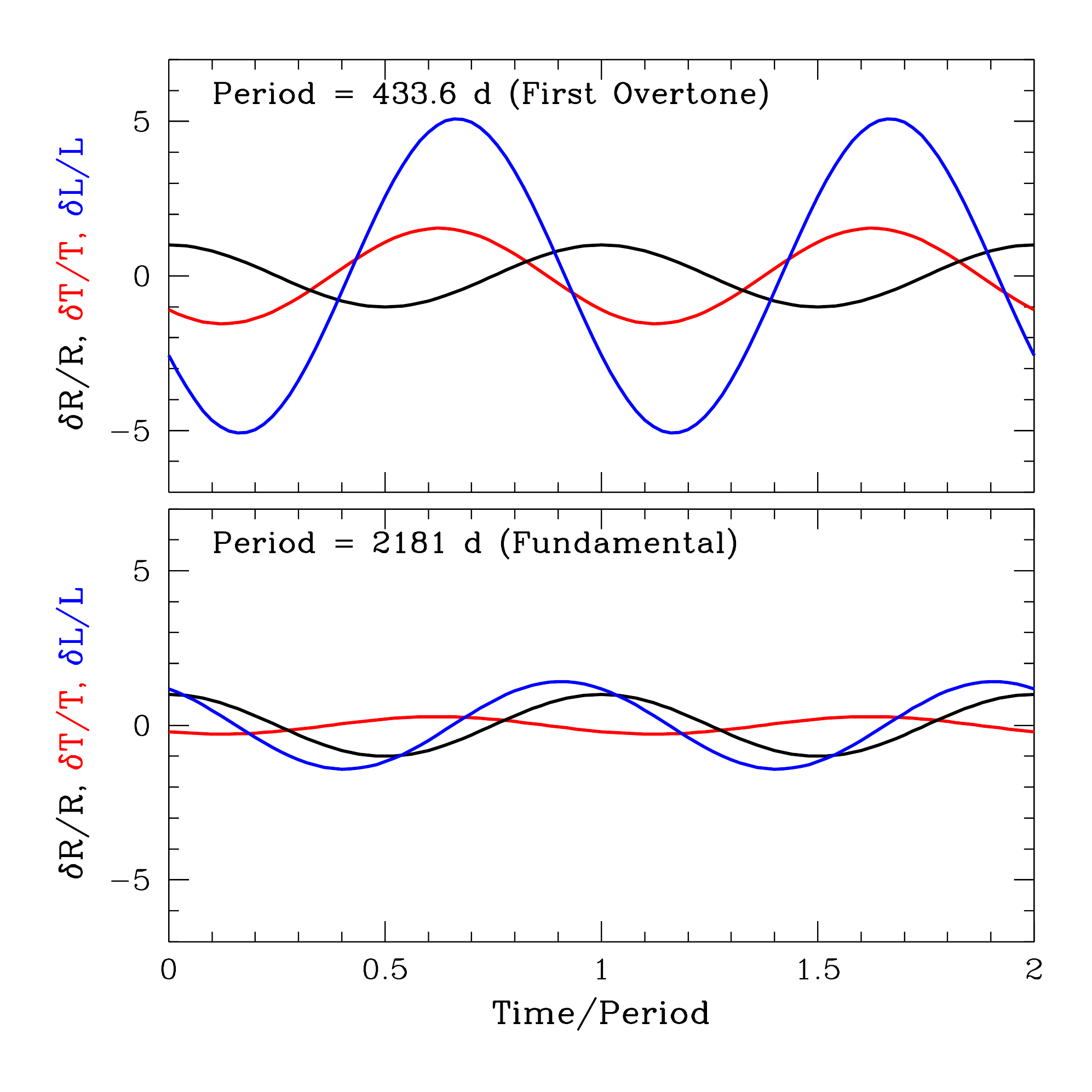}  
\caption{Periodic variations for $\delta R/R$ (black), $\delta T/T$ (red) and
$\delta L/L$ (blue) at the surface, predicted by the linear perturbation analysis 
for the two lowest radial modes excited in model\,C (Table\,\ref{tab:models}),
while growth of amplitude is not included.
The lower and the upper panels are for the fundamental ($P_1$) and 
first overtone ($P_2$) modes
 with periods of 2181\,d and 433.6\,d, respectively. 
The amplitudes are arbitrarily normalized as $\delta R/R=1$ at $t=0$ for both
cases.}
\label{fig:LCmodel}
\end{figure}
Figure\,\ref{fig:LCmodel} shows theoretical periodic variations of fractional
radius (black), temperature (red), and luminosity (blue) at the surface 
for the radial fundamental mode $P_1$ (2181\,d; lower panel) and the first overtone 
$P_2$ (433.6\,d; upper panel) excited in model\,C (Table\,\ref{tab:models}).
The amplitudes are normalized as $\delta R/R=1$ at $t=0$ for both cases and 
fixed by suppressing a possible amplitude growth with time.

The amplitudes of the luminosity and the temperature variations of 
the $P_1$ mode are much smaller than the corresponding amplitudes of the $P_2$ mode.
In addition,  the luminosity maximum of $P_1$ mode occurs around the phase 
of maximum displacement $\delta R/R$, while for $P_2$ mode it occurs 
around minimum $\delta R/R$ when $\delta T/T$ is nearly maximum.
The difference comes from the different strength of the non-adiabatic effects, which
 are stronger in the long-period $P_1$ mode than in the $P_2$ mode.
Strong non-adiabaticity reduces the temperature variation significantly
so that the radius effect in the luminosity variation
\begin{equation}
\delta L/L = 4\delta T_{\rm eff}/T_{\rm eff} + 2\delta R/R
\end{equation}
exceeds the temperature effect. 
This explains why the luminosity maximum of the $P_1$ mode occurs around the maximum 
displacement $\delta R/R$.  
In contrast, for the shorter-period $P_2$ mode, 
the temperature perturbation to be large enough 
to exceed the effect of $\delta R/R$ effect so that the luminosity maximum 
occurs near the phase of temperature maximum.

\subsection{Synthesis of semi-periodic variation}

\begin{figure}
\includegraphics[width=\columnwidth]{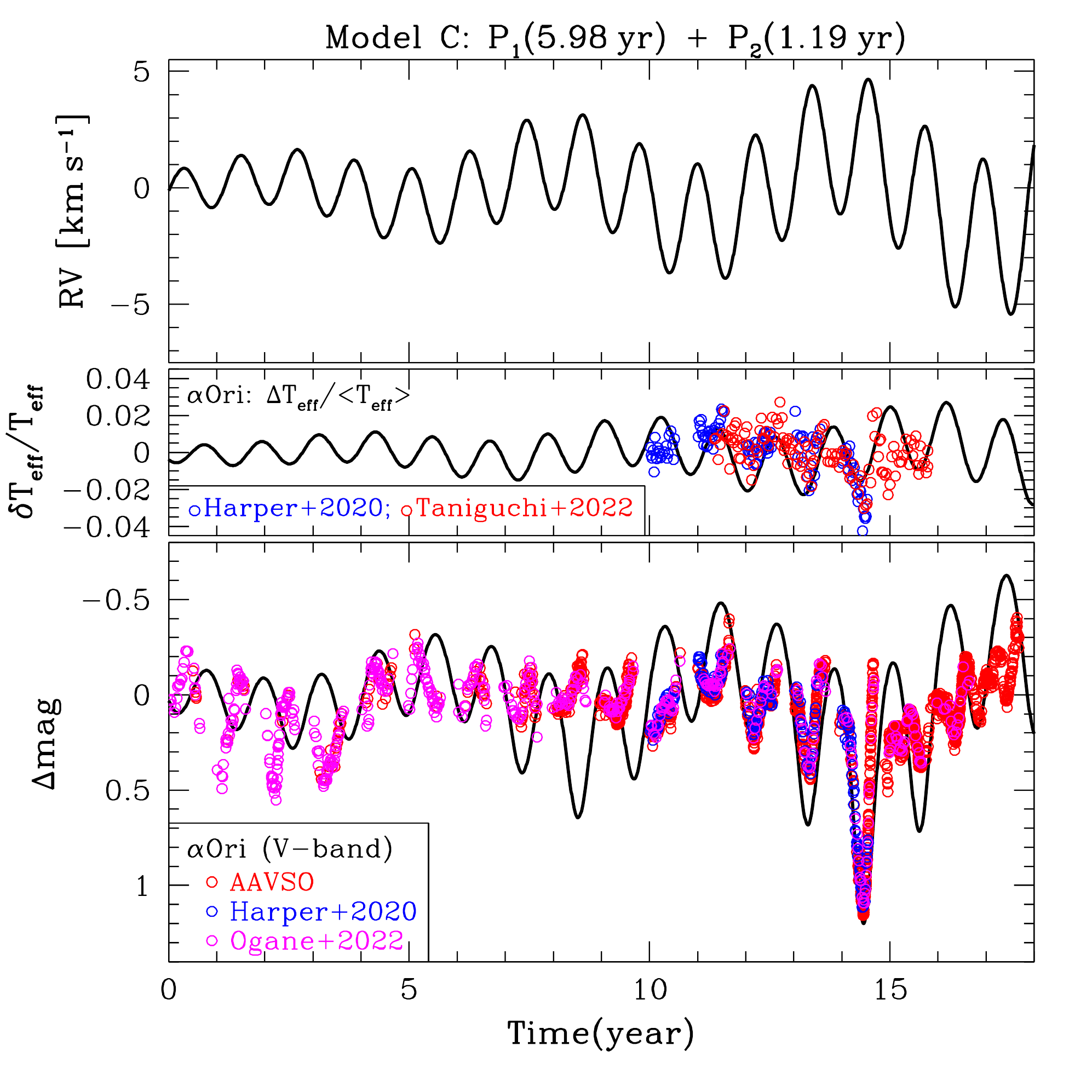} 
\caption{
Radial velocity (RV;top), $T_{\rm eff}$ variation (middle) 
and brightness variation (bottom) versus time.
The solid line in each panel is obtained by combining the fundamental ($P_1$) 
and the first-overtone ($P_2$) radial pulsations excited in 
model C (Table\,\ref{tab:models}). 
The radial velocity in the top panel is set to be positive for 'red-shift'. 
The brightness variations in the bottom panel
is normalized such that the maximum magnitude variation in this time range to be
1.2\, mag, which also scales the amplitudes of model curves
for RV and $T_{\rm eff}$ variations.
The V-band magnitudes of Betelgeuse from AAVSO,
 \citet{Harper2020} and \citet{Ogane2022} are over-plotted, 
where the mean magnitude is assumed
to be 0.5\,mag and the observed dates are shifted by $-2005.6$\,yr to fit with
the model time.   
In the middle panel, relative deviations of $T_{\rm eff}$ from the mean
$\langle T_{\rm eff}\rangle$ of Betelgeuse
obtained by \citet{Harper2020} and \citet{Taniguchi2022} are also plotted.  
 }
\label{fig:radVmag}
\end{figure}

All recent period analyses by various authors for the Betelgeuse light curves 
yielded at least periods $P_1$ and $P_2$ (see e.g. Table\,\ref{tab:periods}).
This implies that the main features of the light variations of Betelgeuse
should be qualitatively represented by a superposition of the periodic
variations of $P_1$ and $P_2$.  
In order to confirm the property, we have constructed a synthetic semi-periodic 
variation by adding the two pulsation modes (Figure\,\ref{fig:radVmag}).
We have assumed arbitrarily that the two modes start (at $t=0$) 
with the same amplitudes to each other.
While a linear pulsation analysis assumes the (infinitesimal) amplitude 
proportional to $\exp(-\sigma_it)$ with the imaginary part of 
the eigenfrequency $\sigma_i$ ($<0$ for an excited mode), we assume 
a slower growth as $(1 - \sigma_i t)$ because a rapid growth should be 
suppressed by non-linear effects for the (observable) finite amplitude pulsation.
The radial velocity, RV is calculated as $-(2/3)\delta R/dt$, where a factor of 2/3
presents the projection effect from the spherical surface.
The RV variation (top panel) is plotted as red-shift (contraction) upward, and
the luminosity variation is plotted in magnitude,
$\Delta{\rm mag}\equiv-2.5\log(1+\delta L/L)$, where $\delta R/R$ 
and $\delta L/L$ are sums of the two pulsation modes. 

The $1.2$-yr ($P_2$; first overtone) mode variations are modulated by the superposition
of the $6.0$-yr ($P_1$; fundamental) mode pulsation.  
The modulation amplitude grows with time as pulsation amplitudes grow.
It is particularly interesting that the brightness variation 
(bottom panel of Fig.\,\ref{fig:radVmag})
has a deep minimum at $t\approx14.5$ mimicking the Great Dimming of Betelgeuse.
To fit the luminosity minimum with the Great Dimming
\footnote{Although complicated phenomena such as dark spots and dust formation occurred during the Great Dimming as summarized in \citet{Wheeler2023}, 
here for simplicity, we have taken into account only pulsational variations for our qualitative synthetic light curve.}  
we have normalized the luminosity variation such that 
$\Delta{\rm mag} = 1.2$\,mag at the
minimum ($t=14.5$) by multiplying a factor to $\delta L(t)$. 
RV(t) and $\delta T_{\rm eff}(t)$ are scaled to be consistent with
the luminosity variation. 
On the synthesized light curve, we have superposed V-band photometry
 results of Betelgeuse from various sources (indicated in this figure), in which all
observed magnitudes are shifted by $-0.5$\,mag (assuming the mean magnitude to
be $0.5$\,mag), and observed time is shifted by $-2005.6$\,yr to fit
the time of the theoretical minimum (at 14.5\,yr) with 
the Great Dimming at $2020.1$\,yr. 

During the two cycles of $P_2$ prior to the Great Dimming, the synthesized light curve
agrees with observation showing minima getting deeper, 
which seems to indicate the Great Dimming to be caused partially
by a constructive interference between the fundamental and the first overtone
pulsations \citep{Guinan2019,Harper2020}.  
Long before the Great Dimming the low-amplitude modulations 
associated with the fundamental mode ($P_1$) approximately trace the envelope
of local maxima of the light variation. 
However, the phases of observational $P_2$ pulsation deviate considerably from the synthetic curve. 
This might mean that a phase modulation occurred  in the first overtone ($P_2$) 
pulsation, or shorter period variations of $P_3$ and $P_4$,  
not included in our synthetic light curve, significantly modify the first overtone pulsation.

Just after the Great Dimming, the $P_2$\,($434$-d) pulsation suddenly became invisible  
and is replaced with low-amplitude shorter timescale $\sim200$-d variations
 \citep{Dupree2022,Jadlovsky2023}, while the fundamental mode ($P_1$) pulsation
seems to remain intact.
This disappearance of the first-overtone pulsation may 
suggest that a massive mass loss \citep{Jadlovsky2023,Dupree2022} 
at the Great Dimming disturbed significantly the $P_2$ pulsation.
Since the growth time of the first-overtone  
pulsation is about 5.4\,yr (Fig.\,\ref{fig:xiwork}), 
the $434$-d pulsation may appear again around 2025.
\citet{MacLeod2023} also predicts the re-appearance of the $\sim400$-d pulsation 
in a similar timescale (5--10 years) although they regard it as fundamental mode.

The predicted effective temperature variation (middle panel of
 Figure\,\ref{fig:radVmag}) hardly  modulates with the period $P_1$. 
This is understood as that the temperature
variation of the fundamental mode (corresponding to $P_1$) is suppressed 
significantly due to the
strong nonadiabatic effect as seen in Figure\,\ref{fig:LCmodel}.
Effective temperature variations of Betelgeuse obtained by
\citet{Harper2020} and \citet{Taniguchi2022} are plotted for comparison.
While observed results have some wiggles that are probably caused by shorter 
period pulsations not considered here, the observed range of $T_{\rm eff}$ 
variations are comparable with our simple two-mode prediction.   
We note that \citet{Wasatonic2022} and \citet{Mittag2023} also obtained similar variations in the
effective temperature of Betelgeuse.   

Around the Great Dimming, the synthesized RV curve 
(top panel) attains maximum (red-shift) 0.1-yr after the minimum brightness.
This phase relation agrees with the observed relation \citep{Dupree2022,MacLeod2023}.
Our two-mode synthesis predicts $4.5$\,km\,s$^{-1}$ at the maximum, which is
comparable with observed values \citep{Kravchenko2021,Dupree2022}.
However, the observed minimum RV (maximum expansion velocity) 
about $-6$\,km\,s$^{-1}$ that occurred $\sim$200-d before the Great Dimming 
is smaller than our model prediction $-1$\,km\,s$^{-1}$.
The discrepancy may be consistent with the emergence of a shock 
in the expansion phase found by \citet{Kravchenko2021} and \citet{Dupree2022},
or a breakout of a convective plume as discussed in \citet{MacLeod2023}. 
While various complex phenomena are involved in the RV variations,
the diameter of Betelgeuse seems to change by $5\sim10\%$ on time-scales comparable
to pulsation periods \citep{Townes2009,Taniguchi2022}.

\subsection{Surface CNO abundance}  

\begin{figure}
\includegraphics[width=\columnwidth]{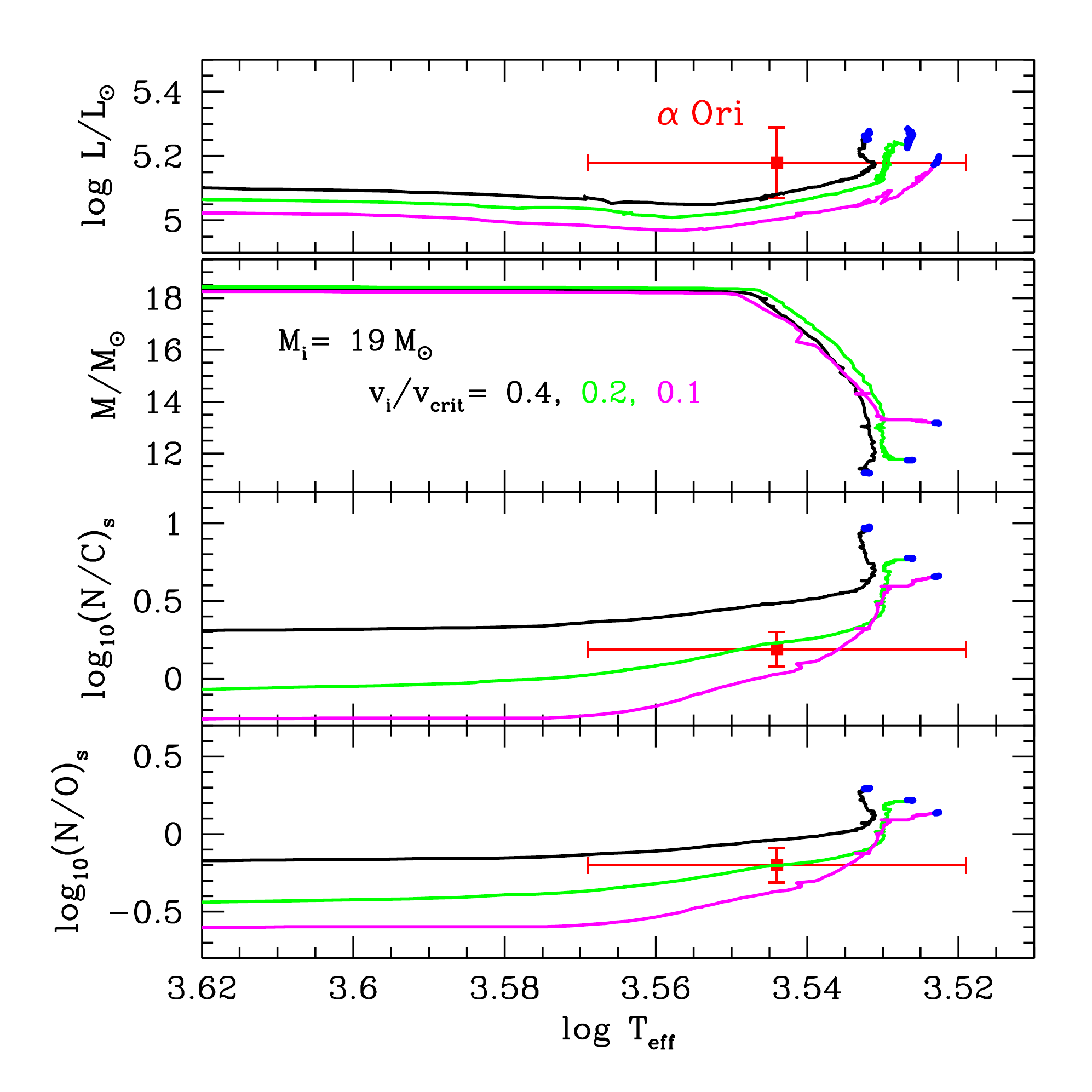}  
\caption{Evolutions of luminosity, mass, and surface CNO ratios 
for an initial mass of $19\,M_\odot$ are compared with the observational data
of Betelgeuse ($\alpha$\,Ori). 
Various cases of the initial rotation frequency versus critical one 
$\varv_{\rm i}/\varv_{\rm crit}$ are color coded as shown 
in the second panel from the top.
The core carbon burning stage is indicated by blue color.
The abundance ratios N/C and N/O stand for the number ratios, while observed 
ones are adopted from \citet{Carr2000}. }
\label{fig:cno}
\end{figure}

Figure\,\ref{fig:cno} shows the evolutions of luminosity, mass, 
and CNO abundance ratios at the surface for 
models with different initial (at ZAMS) rotation velocity given as 
$\varv_{\rm i}/\varv_{\rm crit}=0.4,0.2,0.1$ ($M_i=19\,M_\odot$).
The second panel from the top shows that the mass-loss occurs mainly in the
red-supergiant range (core He burning stage), where up to $~6\,M_\odot$ is lost.
The lost mass is larger for a larger $\varv_{\rm i}$ because of 
a larger luminosity associated with larger helium core, which was 
produced by extensive rotational mixing around the convective core during the
main-sequence evolution.

Because of the large mass loss in the red-supergiant stage, CNO processed
matter emerges to the surface and increases N/C and N/O ratios.
The N/C and N/O ratios of the models in the core carbon burning stage, 
whose pulsation periods agree well with the observed periods of Betelgeuse, 
are, unfortunately, larger than the observed values obtained by \citet{Carr2000}.
As a numerical example, the model predicts a N/C that is 0.5 dex
larger than measured.
This discrepancy seems to indicate that the rotational diffusive mixing is 
overestimated.
The present computation has been performed using specific choices for the
diffusion coefficients describing the transport by shear turbulence and meridional currents \citep[see references in][]{Ekstroem2012}. The choice made in the present computation actually favors the mixing.
Other choices would have produced smaller results. As a numerical example, models of 15 M$_\odot$ with an initial rotation equal to 40\% the critical velocity at solar metallicity have been computed with different choices of these diffusion coefficients (see Nandal et al.2023 in preparation). The N/C ratio at the surface when 
$\log T_{\rm eff} =3.6$  can be reduced up to 0.5 dex changing the expressions of the diffusion coefficients describing the shear mixing. Thus indeed, the solution to that problem can be linked to the specific physics used here for rotational mixing. 
These models would however present similar structure of their envelope at the red supergiant stage and thus we do not expect this will influence the properties of the pulsation modes during that stage.

\section{Discussion}
Our models that excite four radial pulsations consistent with 
observed four periods of Betelgeuse (Table\,\ref{tab:models}) have radii larger 
than $1200\,R_\odot$, which are much larger than the previous 
estimates based on 
seismological predictions ($\sim700$ to $\sim 900\,R_\odot$) 
by \citet{Joyce2020} and \citet{Dolan2016}. 
The difference arises from the fact that the previous analysis fitted 
the period $P_2$ ($\sim420$\,d) with the period of the radial fundamental 
pulsation, regarding the periodicity 
$P_1$ ($\sim2200$\,d) to be a non-pulsational origin.   
In contrast, we have fitted the period $P_1$ with the fundamental non-adiabatic
pulsation period of a core-carbon-burning model having a radius of 
$\sim1300\,R_\odot$ (Table\,\ref{tab:models}).
The model excites not only the fundamental mode which fits with $P_1$ but also
excites the first, second, and third overtones whose periods agree with
the observed periods $P_2$, $P_3$, and $P_4$, respectively.

The large radii of our models have some supports
from some of the interferometric observations of the angular diameter,
although the determinations are affected by various complex factors \citep{Dolan2016}.
\citet{Haubois2009} obtained the Rosseland angular diameter of Betelgeuse to be
$45.03 \pm 0.12$\,mas which was confirmed by \citet{Neilson2011}.
Combining the angular diameter with the distance $222^{+48}_{-34}\,$pc 
\citep{Harper2017} yields $R(\alpha\,{\rm Ori}) = 1074^{+232}_{-165}\,R_\odot$,
which is consistent with our models in Table\,\ref{tab:models}.
Furthermore, \citet{Cannon2023} obtained an angular diameter of $59.02\pm0.64$\,mas   
in the spectral range between 8 and $8.75\,\mu$m, which 
corresponds to a radius $1409^{+319}_{-229}R_\odot$ at the same distance as above,
while it is known that measurements with longer wavelength 
yield larger angular diameters.
{\footnote{\citet{Molnar2023} claims that the radius of Betelgeuse 
should be limited to $<1100\,R_\odot$. However, we do not consider the
limit to be seriously discrepant to the \citet{Haubois2009}'s result, because 
various uncertainties would enter in the 
measurements of the angular diameter.}  }

Furthermore, fitting interferometric observations with the limb-darkening law,
\citet{Neilson2011} obtained the radius to mass ratio 
$R/M = 82.17^{+13.32}_{-11.51}R_\odot/M_\odot$, while our models 
in Table\,\ref{tab:models} give $R/M = 110 \pm 2\,R_\odot/M_\odot$.
Although $R/M$ of our models are slightly larger than the \citet{Neilson2011}'s estimate, we do not think the difference serious because
the mass-loss rate assumed as a function of stellar parameters certainly has 
some deviations from the real mass-loss rates.
\citet{Lobel2000} obtained a surface gravity of $\log g=-0.5$
from metal absorption lines of Betelgeuse, 
while our models in Table\ref{tab:models} have $\log g= -0.7$.
We consider that the difference is not serious, because a typical uncertainty
in the measurements of $\log g$ is $\pm0.2$ \citep[e.g.][]{Smalley2005}. 

\citet{Kervella2018} estimated an equatorial rotation velocity of
$\varv_{\rm eq}=6.5^{+4.1}_{-0.8}$km\,s$^{-1}$.  
In contrast, our models of Betelgeuse have rotation 
velocity less than 0.1\,km\,s$^{-1}$ 
even if they were rotating at 40\% of critical value at the ZAMS stage. 
The lack of rotation is common to the single-star models of Betelgeuse, as \citet{Wheeler2017} proposes a merger with another star to get additional angular momentum  \citep[see also][]{Sullivan2020}. 

Recently, \citet{Neuhauser2022} claimed to find evidences in pre-telescopic records suggesting that the color of Betelgeuse was yellow 2000 yr ago. 
Based on the evolution models by \citet{Choi2016}, they concluded that 
Betelgeuse having an initial mass of $\sim14M_\odot$
crossed the Herzsprung gap 1000 years ago and settled in the early stage of He burning. Their interpretation contradicts with our carbon-burning models.    
However, we do not consider their interpretation unique.
We discuss here another possibility associated with a blue loop 
from the red supergiant branch.
The surface-temperature evolution of supergiants is very sensitive to the interior chemical composition gradients, envelope mass (or mass-loss rates), core mass, etc.
A small or large blue loop can occur depending on subtle differences in these
parameters as shown in \citet{Georgy2012} and \citet{Meynet2013,Meynet2015}.
These evolutions are not well understood theoretically. 
Observationally, it is known that the progenitors of some Type II supernovae were yellow supergiants \citep{Georgy2012,Smartt2015}, which indicates blue loops to actually occur even in the very late evolutionary stage. 
Therefore, we interpret \citet{Neuhauser2022}'s finding 
as that Betelgeuse returned recently from such a blue-loop evolution. 
Although our models do not have blue loops, we may be able to have a blue loop 
during the red-supergiant evolution by tuning parameters, which we leave for the future
research.

\begin{figure}
\includegraphics[width=\columnwidth]{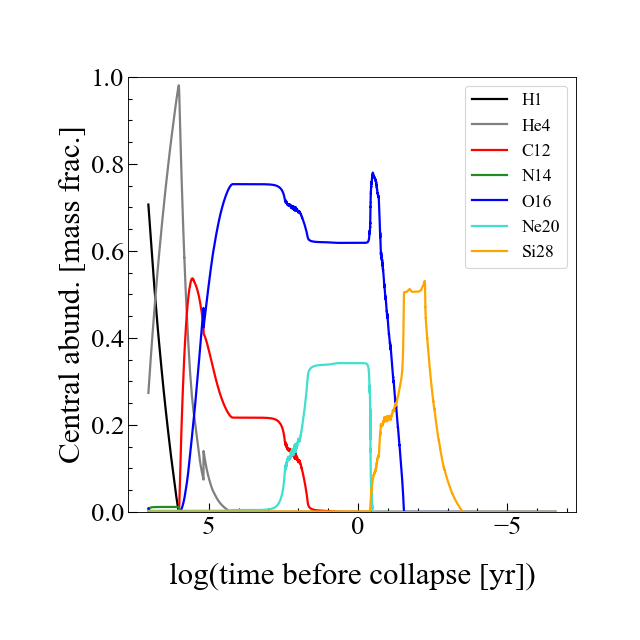}  
\caption{Central abundances of various elements versus time (in logarithm of base 10)
 to collapse for the model of $M_{\rm i}=19\,M_\odot$ with the initial rotation 
velocity $\varv_{\rm i}=0.2\,\varv_{\rm crit}$. }
\label{fig:cent_abun}
\end{figure}

We have found that four periods of the photometric variations of Betelgeuse
are consistent with the four lowest order radial pulsations in the carbon burning
models which evolved from an initial stellar mass of $19\,M_\odot$.
As shown in Table\,\ref{tab:models}, the evolution stage can be very close to 
the carbon exhaustion (as models A, B, and C). 
In fact, it is not possible to determine the exact evolutionary stage, because
surface conditions hardly change in the late stage close to the carbon exhaustion
and beyond, while the property of radial pulsation depends only in the envelope
structure. 
To see the time to a core-collapse after the core carbon exhaustion, we have
extended the evolution for the model with $\varv_{\rm i} = 0.2\varv_{\rm crit}$ to
the silicon exhaustion. 
Figure\,\ref{fig:cent_abun} depicts the central abundance of
elements of the model versus time to the collapse.
In model D, the central carbon would be exhausted in about 260 years, while less time is needed in the other models in Table\,\ref{tab:models}.
After the central carbon exhaustion, Figure\,\ref{fig:cent_abun} indicates that 
the core will collapse in a few tens of years.
This suggests 
Betelgeuse to be a good candidate for the next Galactic 
supernova which occurs very near to us.

Recently, the supernova SN2023ixf is discovered in the nearby galaxy M101 \citep{Itagaki2023}. Interestingly, the pre-explosion observation archives show that the progenitor of the supernova was a pulsating red supergiant, for which \citet{Jencson2023} obtained the period $1119.4^{+132.4}_{-233.3}$ days and the luminosity $\log L/L_\odot = 5.1\pm 0.2$ \citep[similar period and luminosity were also obtained by][]{Soraisam2023}. For comparison, the period range is shown in Figure\,\ref{fig:te_lum_peri} by a horizontal gray line. The position in the PL relation suggests that the progenitor of SN2023ixt might be slightly less massive compared with Betelgeuse, which is consistent with the initial mass $17\pm4\,M_\odot$ estimated by \citet{Jencson2023}.

\section{Conclusion}
We have found carbon-burning models that excite the radial fundamental mode, 
as well as the first, second, and third overtones. 
The periods excited pulsation modes agree with 
periods of 2190, 417, 230, and 185\,d that had been detected in Betelgeuse.  
On the HR diagram, these models are located within the allowed range of 
effective temperature and luminosity of Betelgeuse.
Beginning with a mass of $19\,M_\odot$ at ZAMS 
(with a rotation velocity of 0.2 or 0.4\,$\varv_{\rm crit}$),  
the models lose significant mass mainly in the core-He burning stage 
to have a mass of $11\sim12\,M_\odot$ in the core carbon-burning stage.  
A large radius of about $1300\,R_\odot$ 
(needed for the long-period fundamental mode) is supported
by some interferometric measurements of the angular diameter 
combined with the distance $222^{+48}_{-34}$\,pc\,\citep{Harper2017}.
We conclude that according to our seismic and evolutionary models
Betelgeuse is likely in a late phase (or near the end) of the core carbon burning.
After carbon is exhausted (likely in less than $\sim300$\,years) in the core, a core-collapse leading to a supernova explosion is expected in a few tens of years.

\section*{Acknowledgements}
The authors thank Dr Nami Mowlavi for interesting discussions 
on the identification of the pulsation modes of Betelgeuse.
We also thank Daisuke Taniguchi for allowing us to use the 
$T_{\rm eff}$ data presented in \citet{Taniguchi2022}. 
We acknowledge with thanks the variable star observations from the AAVSO International Database contributed by observers worldwide and used in this research.
DN, GM, and SE have received funding from the European Research Council (ERC) 
under the European Union's Horizon 2020 research and innovation program 
(grant agreement No 833925, project STAREX).

\section*{Data Availability}
The data underlying this article and pulsation code will be shared on reasonable 
request to the authors, while some interactive instructions 
how to use it would be needed for the latter.



\bibliographystyle{mnras}
\bibliography{alpOri} 

\begin{thebibliography}{}
\makeatletter
\relax
\def\mn@urlcharsother{\let\do\@makeother \do\$\do\&\do\#\do\^\do\_\do\%\do\~}
\def\mn@doi{\begingroup\mn@urlcharsother \@ifnextchar [ {\mn@doi@}
  {\mn@doi@[]}}
\def\mn@doi@[#1]#2{\def\@tempa{#1}\ifx\@tempa\@empty \href
  {http://dx.doi.org/#2} {doi:#2}\else \href {http://dx.doi.org/#2} {#1}\fi
  \endgroup}
\def\mn@eprint#1#2{\mn@eprint@#1:#2::\@nil}
\def\mn@eprint@arXiv#1{\href {http://arxiv.org/abs/#1} {{\tt arXiv:#1}}}
\def\mn@eprint@dblp#1{\href {http://dblp.uni-trier.de/rec/bibtex/#1.xml}
  {dblp:#1}}
\def\mn@eprint@#1:#2:#3:#4\@nil{\def\@tempa {#1}\def\@tempb {#2}\def\@tempc
  {#3}\ifx \@tempc \@empty \let \@tempc \@tempb \let \@tempb \@tempa \fi \ifx
  \@tempb \@empty \def\@tempb {arXiv}\fi \@ifundefined
  {mn@eprint@\@tempb}{\@tempb:\@tempc}{\expandafter \expandafter \csname
  mn@eprint@\@tempb\endcsname \expandafter{\@tempc}}}

\bibitem[\protect\citeauthoryear{{Anderson}, {Saio}, {Ekstr{\"o}m}, {Georgy}
  \& {Meynet}}{{Anderson} et~al.}{2016}]{Anderson2016}
{Anderson} R.~I.,  {Saio} H.,  {Ekstr{\"o}m} S.,  {Georgy} C.,   {Meynet} G.,
  2016, \mn@doi [\aap] {10.1051/0004-6361/201528031}, \href
  {https://ui.adsabs.harvard.edu/abs/2016A&A...591A...8A} {591, A8}

\bibitem[\protect\citeauthoryear{{Cannon} et~al.,}{{Cannon}
  et~al.}{2023}]{Cannon2023}
{Cannon} E.,  et~al., 2023, \mn@doi [arXiv e-prints]
  {10.48550/arXiv.2303.08892}, \href
  {https://ui.adsabs.harvard.edu/abs/2023arXiv230308892C} {p. arXiv:2303.08892}

\bibitem[\protect\citeauthoryear{{Carr}, {Sellgren}  \& {Balachandran}}{{Carr}
  et~al.}{2000}]{Carr2000}
{Carr} J.~S.,  {Sellgren} K.,   {Balachandran} S.~C.,  2000, \mn@doi [\apj]
  {10.1086/308340}, \href
  {https://ui.adsabs.harvard.edu/abs/2000ApJ...530..307C} {530, 307}

\bibitem[\protect\citeauthoryear{{Chatys}, {Bedding}, {Murphy}, {Kiss}, {Dobie}
   \& {Grindlay}}{{Chatys} et~al.}{2019}]{Chatys2019}
{Chatys} F.~W.,  {Bedding} T.~R.,  {Murphy} S.~J.,  {Kiss} L.~L.,  {Dobie} D.,
   {Grindlay} J.~E.,  2019, \mn@doi [\mnras] {10.1093/mnras/stz1584}, \href
  {https://ui.adsabs.harvard.edu/abs/2019MNRAS.487.4832C} {487, 4832}

\bibitem[\protect\citeauthoryear{{Choi}, {Dotter}, {Conroy}, {Cantiello},
  {Paxton}  \& {Johnson}}{{Choi} et~al.}{2016}]{Choi2016}
{Choi} J.,  {Dotter} A.,  {Conroy} C.,  {Cantiello} M.,  {Paxton} B.,
  {Johnson} B.~D.,  2016, \mn@doi [\apj] {10.3847/0004-637X/823/2/102}, \href
  {https://ui.adsabs.harvard.edu/abs/2016ApJ...823..102C} {823, 102}

\bibitem[\protect\citeauthoryear{{Cutri} et~al.,}{{Cutri}
  et~al.}{2003}]{Cutri2003}
{Cutri} R.~M.,  et~al., 2003, VizieR Online Data Catalog, \href
  {https://ui.adsabs.harvard.edu/abs/2003yCat.2246....0C} {p. II/246}

\bibitem[\protect\citeauthoryear{{Dharmawardena}, {Mairs}, {Scicluna}, {Bell},
  {McDonald}, {Menten}, {Weiss}  \& {Zijlstra}}{{Dharmawardena}
  et~al.}{2020}]{Dharmawardena2020}
{Dharmawardena} T.~E.,  {Mairs} S.,  {Scicluna} P.,  {Bell} G.,  {McDonald} I.,
   {Menten} K.,  {Weiss} A.,   {Zijlstra} A.,  2020, \mn@doi [\apjl]
  {10.3847/2041-8213/ab9ca6}, \href
  {https://ui.adsabs.harvard.edu/abs/2020ApJ...897L...9D} {897, L9}

\bibitem[\protect\citeauthoryear{{Dolan}, {Mathews}, {Lam}, {Quynh Lan},
  {Herczeg}  \& {Dearborn}}{{Dolan} et~al.}{2016}]{Dolan2016}
{Dolan} M.~M.,  {Mathews} G.~J.,  {Lam} D.~D.,  {Quynh Lan} N.,  {Herczeg}
  G.~J.,   {Dearborn} D. S.~P.,  2016, \mn@doi [\apj]
  {10.3847/0004-637X/819/1/7}, \href
  {https://ui.adsabs.harvard.edu/abs/2016ApJ...819....7D} {819, 7}

\bibitem[\protect\citeauthoryear{{Dupree} et~al.,}{{Dupree}
  et~al.}{2022}]{Dupree2022}
{Dupree} A.~K.,  et~al., 2022, \mn@doi [\apj] {10.3847/1538-4357/ac7853}, \href
  {https://ui.adsabs.harvard.edu/abs/2022ApJ...936...18D} {936, 18}

\bibitem[\protect\citeauthoryear{{Ekstr{\"o}m} et~al.,}{{Ekstr{\"o}m}
  et~al.}{2012}]{Ekstroem2012}
{Ekstr{\"o}m} S.,  et~al., 2012, \mn@doi [\aap] {10.1051/0004-6361/201117751},
  \href {https://ui.adsabs.harvard.edu/abs/2012A&A...537A.146E} {537, A146}

\bibitem[\protect\citeauthoryear{{Gabriel}, {Scuflaire}, {Noels}  \&
  {Boury}}{{Gabriel} et~al.}{1975}]{Gabriel1975}
{Gabriel} M.,  {Scuflaire} R.,  {Noels} A.,   {Boury} A.,  1975, \aap, \href
  {https://ui.adsabs.harvard.edu/abs/1975A&A....40...33G} {40, 33}

\bibitem[\protect\citeauthoryear{{Georgy}}{{Georgy}}{2012}]{Georgy2012}
{Georgy} C.,  2012, \mn@doi [\aap] {10.1051/0004-6361/201118372}, \href
  {https://ui.adsabs.harvard.edu/abs/2012A&A...538L...8G} {538, L8}

\bibitem[\protect\citeauthoryear{{Gonczi} \& {Osaki}}{{Gonczi} \&
  {Osaki}}{1980}]{Gonczi1980}
{Gonczi} G.,  {Osaki} Y.,  1980, \aap, \href
  {https://ui.adsabs.harvard.edu/abs/1980A&A....84..304G} {84, 304}

\bibitem[\protect\citeauthoryear{{Granzer}, {Weber}, {Strassmeier}  \&
  {Dupree}}{{Granzer} et~al.}{2022}]{Granzer2022}
{Granzer} T.,  {Weber} M.,  {Strassmeier} K.~G.,   {Dupree} A.,  2022, in
  Cambridge Workshop on Cool Stars, Stellar Systems, and the Sun. Cambridge
  Workshop on Cool Stars, Stellar Systems, and the Sun.
p.~185, \mn@doi{10.5281/zenodo.7589936}

\bibitem[\protect\citeauthoryear{{Grigahc{\`e}ne}, {Dupret}, {Gabriel},
  {Garrido}  \& {Scuflaire}}{{Grigahc{\`e}ne} et~al.}{2005}]{Liege2005}
{Grigahc{\`e}ne} A.,  {Dupret} M.~A.,  {Gabriel} M.,  {Garrido} R.,
  {Scuflaire} R.,  2005, \mn@doi [\aap] {10.1051/0004-6361:20041816}, \href
  {https://ui.adsabs.harvard.edu/abs/2005A&A...434.1055G} {434, 1055}

\bibitem[\protect\citeauthoryear{{Guinan}, {Wasatonic}  \&
  {Calderwood}}{{Guinan} et~al.}{2019}]{Guinan2019}
{Guinan} E.~F.,  {Wasatonic} R.~J.,   {Calderwood} T.~J.,  2019, The
  Astronomer's Telegram, \href
  {https://ui.adsabs.harvard.edu/abs/2019ATel13365....1G} {13365, 1}

\bibitem[\protect\citeauthoryear{{Harper}, {Brown}, {Guinan}, {O'Gorman},
  {Richards}, {Kervella}  \& {Decin}}{{Harper} et~al.}{2017}]{Harper2017}
{Harper} G.~M.,  {Brown} A.,  {Guinan} E.~F.,  {O'Gorman} E.,  {Richards}
  A.~M.~S.,  {Kervella} P.,   {Decin} L.,  2017, \mn@doi [\aj]
  {10.3847/1538-3881/aa6ff9}, \href
  {https://ui.adsabs.harvard.edu/abs/2017AJ....154...11H} {154, 11}

\bibitem[\protect\citeauthoryear{{Harper}, {Guinan}, {Wasatonic}  \&
  {Ryde}}{{Harper} et~al.}{2020}]{Harper2020}
{Harper} G.~M.,  {Guinan} E.~F.,  {Wasatonic} R.,   {Ryde} N.,  2020, \mn@doi
  [\apj] {10.3847/1538-4357/abc1f0}, \href
  {https://ui.adsabs.harvard.edu/abs/2020ApJ...905...34H} {905, 34}

\bibitem[\protect\citeauthoryear{{Haubois} et~al.,}{{Haubois}
  et~al.}{2009}]{Haubois2009}
{Haubois} X.,  et~al., 2009, \mn@doi [\aap] {10.1051/0004-6361/200912927},
  \href {https://ui.adsabs.harvard.edu/abs/2009A&A...508..923H} {508, 923}

\bibitem[\protect\citeauthoryear{{Henyey}, {Vardya}  \& {Bodenheimer}}{{Henyey}
  et~al.}{1965}]{Henyey1965}
{Henyey} L.,  {Vardya} M.~S.,   {Bodenheimer} P.,  1965, \mn@doi [\apj]
  {10.1086/148357}, \href
  {https://ui.adsabs.harvard.edu/abs/1965ApJ...142..841H} {142, 841}

\bibitem[\protect\citeauthoryear{{Itagaki}}{{Itagaki}}{2023}]{Itagaki2023}
{Itagaki} K.,  2023, Transient Name Server Discovery Report, \href
  {https://ui.adsabs.harvard.edu/abs/2023TNSTR1158....1I} {2023-1158, 1}

\bibitem[\protect\citeauthoryear{{Jadlovsk{\'y}}, {Krti{\v{c}}ka}, {Paunzen}
  \& {{\v{S}}tefl}}{{Jadlovsk{\'y}} et~al.}{2023}]{Jadlovsky2023}
{Jadlovsk{\'y}} D.,  {Krti{\v{c}}ka} J.,  {Paunzen} E.,   {{\v{S}}tefl} V.,
  2023, \mn@doi [\na] {10.1016/j.newast.2022.101962}, \href
  {https://ui.adsabs.harvard.edu/abs/2023NewA...9901962J} {99, 101962}

\bibitem[\protect\citeauthoryear{{Jencson} et~al.,}{{Jencson}
  et~al.}{2023}]{Jencson2023}
{Jencson} J.~E.,  et~al., 2023, \mn@doi [arXiv e-prints]
  {10.48550/arXiv.2306.08678}, \href
  {https://ui.adsabs.harvard.edu/abs/2023arXiv230608678J} {p. arXiv:2306.08678}

\bibitem[\protect\citeauthoryear{{Josselin}, {Blommaert}, {Groenewegen},
  {Omont}  \& {Li}}{{Josselin} et~al.}{2000}]{Josselin2000}
{Josselin} E.,  {Blommaert} J.~A.~D.~L.,  {Groenewegen} M.~A.~T.,  {Omont} A.,
   {Li} F.~L.,  2000, \aap, \href
  {https://ui.adsabs.harvard.edu/abs/2000A&A...357..225J} {357, 225}

\bibitem[\protect\citeauthoryear{{Joyce}, {Leung}, {Moln{\'a}r}, {Ireland},
  {Kobayashi}  \& {Nomoto}}{{Joyce} et~al.}{2020}]{Joyce2020}
{Joyce} M.,  {Leung} S.-C.,  {Moln{\'a}r} L.,  {Ireland} M.,  {Kobayashi} C.,
  {Nomoto} K.,  2020, \mn@doi [\apj] {10.3847/1538-4357/abb8db}, \href
  {https://ui.adsabs.harvard.edu/abs/2020ApJ...902...63J} {902, 63}

\bibitem[\protect\citeauthoryear{{Kervella} et~al.,}{{Kervella}
  et~al.}{2018}]{Kervella2018}
{Kervella} P.,  et~al., 2018, \mn@doi [\aap] {10.1051/0004-6361/201731761},
  \href {https://ui.adsabs.harvard.edu/abs/2018A&A...609A..67K} {609, A67}

\bibitem[\protect\citeauthoryear{{Kiss}, {Szab{\'o}}  \& {Bedding}}{{Kiss}
  et~al.}{2006}]{Kiss2006}
{Kiss} L.~L.,  {Szab{\'o}} G.~M.,   {Bedding} T.~R.,  2006, \mn@doi [\mnras]
  {10.1111/j.1365-2966.2006.10973.x}, \href
  {https://ui.adsabs.harvard.edu/abs/2006MNRAS.372.1721K} {372, 1721}

\bibitem[\protect\citeauthoryear{{Kravchenko} et~al.,}{{Kravchenko}
  et~al.}{2021}]{Kravchenko2021}
{Kravchenko} K.,  et~al., 2021, \mn@doi [\aap] {10.1051/0004-6361/202039801},
  \href {https://ui.adsabs.harvard.edu/abs/2021A&A...650L..17K} {650, L17}

\bibitem[\protect\citeauthoryear{{Levesque} \& {Massey}}{{Levesque} \&
  {Massey}}{2020}]{Levesque2020}
{Levesque} E.~M.,  {Massey} P.,  2020, \mn@doi [\apjl]
  {10.3847/2041-8213/ab7935}, \href
  {https://ui.adsabs.harvard.edu/abs/2020ApJ...891L..37L} {891, L37}

\bibitem[\protect\citeauthoryear{{Levesque}, {Massey}, {Olsen}, {Plez},
  {Josselin}, {Maeder}  \& {Meynet}}{{Levesque} et~al.}{2005}]{Levesque2005}
{Levesque} E.~M.,  {Massey} P.,  {Olsen} K.~A.~G.,  {Plez} B.,  {Josselin} E.,
  {Maeder} A.,   {Meynet} G.,  2005, \mn@doi [\apj] {10.1086/430901}, \href
  {https://ui.adsabs.harvard.edu/abs/2005ApJ...628..973L} {628, 973}

\bibitem[\protect\citeauthoryear{{Lobel} \& {Dupree}}{{Lobel} \&
  {Dupree}}{2000}]{Lobel2000}
{Lobel} A.,  {Dupree} A.~K.,  2000, \mn@doi [\apj] {10.1086/317784}, \href
  {https://ui.adsabs.harvard.edu/abs/2000ApJ...545..454L} {545, 454}

\bibitem[\protect\citeauthoryear{{Luo}, {Umeda}, {Yoshida}  \&
  {Takahashi}}{{Luo} et~al.}{2022}]{Luo2022}
{Luo} T.,  {Umeda} H.,  {Yoshida} T.,   {Takahashi} K.,  2022, \mn@doi [\apj]
  {10.3847/1538-4357/ac4f5f}, \href
  {https://ui.adsabs.harvard.edu/abs/2022ApJ...927..115L} {927, 115}

\bibitem[\protect\citeauthoryear{{MacLeod}, {Antoni}, {Huang}, {Dupree}  \&
  {Loeb}}{{MacLeod} et~al.}{2023}]{MacLeod2023}
{MacLeod} M.,  {Antoni} A.,  {Huang} C.~D.,  {Dupree} A.,   {Loeb} A.,  2023,
  \mn@doi [arXiv e-prints] {10.48550/arXiv.2305.09732}, \href
  {https://ui.adsabs.harvard.edu/abs/2023arXiv230509732M} {p. arXiv:2305.09732}

\bibitem[\protect\citeauthoryear{{Meynet}, {Haemmerl{\'e}}, {Ekstr{\"o}m},
  {Georgy}, {Groh}  \& {Maeder}}{{Meynet} et~al.}{2013}]{Meynet2013}
{Meynet} G.,  {Haemmerl{\'e}} L.,  {Ekstr{\"o}m} S.,  {Georgy} C.,  {Groh} J.,
   {Maeder} A.,  2013, in {Kervella} P.,  {Le Bertre} T.,   {Perrin} G.,  eds,
  EAS Publications Series Vol. 60, EAS Publications Series. pp 17--28
  (\mn@eprint {arXiv} {1303.1339}), \mn@doi{10.1051/eas/1360002}

\bibitem[\protect\citeauthoryear{{Meynet} et~al.,}{{Meynet}
  et~al.}{2015}]{Meynet2015}
{Meynet} G.,  et~al., 2015, \mn@doi [\aap] {10.1051/0004-6361/201424671}, \href
  {https://ui.adsabs.harvard.edu/abs/2015A&A...575A..60M} {575, A60}

\bibitem[\protect\citeauthoryear{{Mittag}, {Schr{\"o}der}, {Perdelwitz}, {Jack}
   \& {Schmitt}}{{Mittag} et~al.}{2023}]{Mittag2023}
{Mittag} M.,  {Schr{\"o}der} K.~P.,  {Perdelwitz} V.,  {Jack} D.,   {Schmitt}
  J.~H.~M.~M.,  2023, \mn@doi [\aap] {10.1051/0004-6361/202244924}, \href
  {https://ui.adsabs.harvard.edu/abs/2023A&A...669A...9M} {669, A9}

\bibitem[\protect\citeauthoryear{{Moln{\'a}r}, {Joyce}  \&
  {Leung}}{{Moln{\'a}r} et~al.}{2023}]{Molnar2023}
{Moln{\'a}r} L.,  {Joyce} M.,   {Leung} S.-C.,  2023, \mn@doi [Research Notes
  of the American Astronomical Society] {10.3847/2515-5172/acdb7a}, \href
  {https://ui.adsabs.harvard.edu/abs/2023RNAAS...7..119M} {7, 119}

\bibitem[\protect\citeauthoryear{{Montarg{\`e}s} et~al.,}{{Montarg{\`e}s}
  et~al.}{2021}]{Montarges2021}
{Montarg{\`e}s} M.,  et~al., 2021, \mn@doi [\nat] {10.1038/s41586-021-03546-8},
  \href {https://ui.adsabs.harvard.edu/abs/2021Natur.594..365M} {594, 365}

\bibitem[\protect\citeauthoryear{{Nance}, {Sullivan}, {Diaz}  \&
  {Wheeler}}{{Nance} et~al.}{2018}]{Nance2018}
{Nance} S.,  {Sullivan} J.~M.,  {Diaz} M.,   {Wheeler} J.~C.,  2018, \mn@doi
  [\mnras] {10.1093/mnras/sty1418}, \href
  {https://ui.adsabs.harvard.edu/abs/2018MNRAS.479..251N} {479, 251}

\bibitem[\protect\citeauthoryear{{Neilson}, {Lester}  \& {Haubois}}{{Neilson}
  et~al.}{2011}]{Neilson2011}
{Neilson} H.~R.,  {Lester} J.~B.,   {Haubois} X.,  2011, in {Qain} S.,  {Leung}
  K.,  {Zhu} L.,   {Kwok} S.,  eds,  Astronomical Society of the Pacific
  Conference Series Vol. 451, 9th Pacific Rim Conference on Stellar
  Astrophysics. p.~117 (\mn@eprint {arXiv} {1109.4562}),
  \mn@doi{10.48550/arXiv.1109.4562}

\bibitem[\protect\citeauthoryear{{Neuh{\"a}user}, {Torres}, {Mugrauer},
  {Neuh{\"a}user}, {Chapman}, {Luge}  \& {Cosci}}{{Neuh{\"a}user}
  et~al.}{2022}]{Neuhauser2022}
{Neuh{\"a}user} R.,  {Torres} G.,  {Mugrauer} M.,  {Neuh{\"a}user} D.~L.,
  {Chapman} J.,  {Luge} D.,   {Cosci} M.,  2022, \mn@doi [\mnras]
  {10.1093/mnras/stac1969}, \href
  {https://ui.adsabs.harvard.edu/abs/2022MNRAS.516..693N} {516, 693}

\bibitem[\protect\citeauthoryear{{Ogane}, {Ohshima}, {Taniguchi}  \&
  {Takanashi}}{{Ogane} et~al.}{2022}]{Ogane2022}
{Ogane} Y.,  {Ohshima} O.,  {Taniguchi} D.,   {Takanashi} N.,  2022, \mn@doi
  [Open European Journal on Variable Stars] {10.5817/OEJV2022-0233}, \href
  {https://ui.adsabs.harvard.edu/abs/2022OEJV..233....1O} {233, 1}

\bibitem[\protect\citeauthoryear{{Saio}}{{Saio}}{1980}]{Saio1980}
{Saio} H.,  1980, \mn@doi [\apj] {10.1086/158275}, \href
  {https://ui.adsabs.harvard.edu/abs/1980ApJ...240..685S} {240, 685}

\bibitem[\protect\citeauthoryear{{Saio} \& {Cox}}{{Saio} \&
  {Cox}}{1980}]{SaioCox1980}
{Saio} H.,  {Cox} J.~P.,  1980, \mn@doi [\apj] {10.1086/157773}, \href
  {https://ui.adsabs.harvard.edu/abs/1980ApJ...236..549S} {236, 549}

\bibitem[\protect\citeauthoryear{{Saio}, {Winget}  \& {Robinson}}{{Saio}
  et~al.}{1983}]{Saio1983}
{Saio} H.,  {Winget} D.~E.,   {Robinson} E.~L.,  1983, \mn@doi [\apj]
  {10.1086/160740}, \href
  {https://ui.adsabs.harvard.edu/abs/1983ApJ...265..982S} {265, 982}

\bibitem[\protect\citeauthoryear{{Smalley}}{{Smalley}}{2005}]{Smalley2005}
{Smalley} B.,  2005, \mn@doi [Memorie della Societa Astronomica Italiana
  Supplementi] {10.48550/arXiv.astro-ph/0509535}, \href
  {https://ui.adsabs.harvard.edu/abs/2005MSAIS...8..130S} {8, 130}

\bibitem[\protect\citeauthoryear{{Smartt}}{{Smartt}}{2015}]{Smartt2015}
{Smartt} S.~J.,  2015, \mn@doi [\pasa] {10.1017/pasa.2015.17}, \href
  {https://ui.adsabs.harvard.edu/abs/2015PASA...32...16S} {32, e016}

\bibitem[\protect\citeauthoryear{{Soraisam} et~al.,}{{Soraisam}
  et~al.}{2023}]{Soraisam2023}
{Soraisam} M.~D.,  et~al., 2023, \mn@doi [arXiv e-prints]
  {10.48550/arXiv.2306.10783}, \href
  {https://ui.adsabs.harvard.edu/abs/2023arXiv230610783S} {p. arXiv:2306.10783}

\bibitem[\protect\citeauthoryear{{Stothers} \& {Leung}}{{Stothers} \&
  {Leung}}{1971}]{Stothers1971}
{Stothers} R.,  {Leung} K.~C.,  1971, \aap, \href
  {https://ui.adsabs.harvard.edu/abs/1971A&A....10..290S} {10, 290}

\bibitem[\protect\citeauthoryear{{Sullivan}, {Nance}  \& {Wheeler}}{{Sullivan}
  et~al.}{2020}]{Sullivan2020}
{Sullivan} J.~M.,  {Nance} S.,   {Wheeler} J.~C.,  2020, \mn@doi [\apj]
  {10.3847/1538-4357/abc3c9}, \href
  {https://ui.adsabs.harvard.edu/abs/2020ApJ...905..128S} {905, 128}

\bibitem[\protect\citeauthoryear{{Taniguchi}, {Yamazaki}  \& {Uno}}{{Taniguchi}
  et~al.}{2022}]{Taniguchi2022}
{Taniguchi} D.,  {Yamazaki} K.,   {Uno} S.,  2022, \mn@doi [Nature Astronomy]
  {10.1038/s41550-022-01680-5}, \href
  {https://ui.adsabs.harvard.edu/abs/2022NatAs...6..930T} {6, 930}

\bibitem[\protect\citeauthoryear{{Townes}, {Wishnow}, {Hale}  \&
  {Walp}}{{Townes} et~al.}{2009}]{Townes2009}
{Townes} C.~H.,  {Wishnow} E.~H.,  {Hale} D.~D.~S.,   {Walp} B.,  2009, \mn@doi
  [\apjl] {10.1088/0004-637X/697/2/L127}, \href
  {https://ui.adsabs.harvard.edu/abs/2009ApJ...697L.127T} {697, L127}

\bibitem[\protect\citeauthoryear{{Trabucchi}, {Wood}, {Montalb{\'a}n},
  {Marigo}, {Pastorelli}  \& {Girardi}}{{Trabucchi}
  et~al.}{2017}]{Trabucchi2017}
{Trabucchi} M.,  {Wood} P.~R.,  {Montalb{\'a}n} J.,  {Marigo} P.,  {Pastorelli}
  G.,   {Girardi} L.,  2017, \mn@doi [\apj] {10.3847/1538-4357/aa8998}, \href
  {https://ui.adsabs.harvard.edu/abs/2017ApJ...847..139T} {847, 139}

\bibitem[\protect\citeauthoryear{{Unno}}{{Unno}}{1967}]{Unno1967}
{Unno} W.,  1967, \pasj, \href
  {https://ui.adsabs.harvard.edu/abs/1967PASJ...19..140U} {19, 140}

\bibitem[\protect\citeauthoryear{{Wasatonic}}{{Wasatonic}}{2022}]{Wasatonic2022}
{Wasatonic} R.,  2022, JAAVSO, \href
  {https://ui.adsabs.harvard.edu/abs/2022JAVSO..50..205W} {50, 205}

\bibitem[\protect\citeauthoryear{{Wheeler} \& {Chatzopoulos}}{{Wheeler} \&
  {Chatzopoulos}}{2023}]{Wheeler2023}
{Wheeler} J.~C.,  {Chatzopoulos} E.,  2023, \mn@doi [arXiv e-prints]
  {10.48550/arXiv.2306.09449}, \href
  {https://ui.adsabs.harvard.edu/abs/2023arXiv230609449W} {p. arXiv:2306.09449}

\bibitem[\protect\citeauthoryear{{Wheeler} et~al.,}{{Wheeler}
  et~al.}{2017}]{Wheeler2017}
{Wheeler} J.~C.,  et~al., 2017, \mn@doi [\mnras] {10.1093/mnras/stw2893}, \href
  {https://ui.adsabs.harvard.edu/abs/2017MNRAS.465.2654W} {465, 2654}

\bibitem[\protect\citeauthoryear{{Yusof} et~al.,}{{Yusof}
  et~al.}{2022}]{Yusof2022}
{Yusof} N.,  et~al., 2022, \mn@doi [\mnras] {10.1093/mnras/stac230}, \href
  {https://ui.adsabs.harvard.edu/abs/2022MNRAS.511.2814Y} {511, 2814}

\makeatother
\end{thebibliography}



\appendix

\section{Radial pulsation coupled with convection}
In this appendix we derive differential equations 
to solve radial pulsations taking into account 
the coupling with convection based on the mixing-length theory.
In the beginning we follow \citet{Unno1967}'s discussion but later we include
additional terms to avoid rapid oscillations where the turnover time is
longer than the pulsation period \citep[e.g.][]{Liege2005}. 
We summarise differential equations of radial pulsations in the last part 
of this appendix.
 
\subsection{Basic Equations} 
Equations of motion and mass conservation, including convective motion may be written as
\begin{equation}
(\rho+\rho')\left[{\partial\over\partial t}+(\bm{u}+\bm{V}_{\rm c})\cdot\nabla\right]
(\bm{u}+\bm{V}_{\rm c}) = -\nabla (P+P') - (\rho+\rho')\nabla\psi
\label{eq:mom0}
\end{equation}
and
\begin{equation}
{\partial(\rho+\rho')\over\partial t} + \nabla\cdot[(\rho+\rho')(\bm{u}+\bm{V}{\rm_c})]=0,
\label{eq:mass0}
\end{equation}
where $\bm{u}$ and $\bm{V}_{\rm c}$ are pulsation and convection velocities, respectively,
and a prime $'$ such as $\rho',P', T',etc$ means the convective Eulerian perturbation. 
We neglect convective perturbation of the gravitational potential $\psi$.

Conservation of thermal energy may be written as
\begin{equation}
\begin{array}{l}\displaystyle
(\rho+\rho')(T+T')\left[{\partial\over\partial t} + (\bm{u}+\bm{V}_{\rm c})\cdot\nabla\right](S+S')  \cr\displaystyle
\qquad \qquad \qquad = \rho\epsilon_{\rm n} + (\rho\epsilon_{\rm n})' 
- \nabla\cdot(\bm{F}_{\rm rad} + \bm{F}_{\rm rad}'),
\end{array}
\label{eq:energy0}
\end{equation}
where $S$ and $\epsilon_{\rm n}$ are the entropy and the nuclear energy 
generation rate per unit mass, respectively, 
and $\bm{F}_{\rm rad}$ is the radiative energy flux. 

\subsection{Equations for mean fluid}
Taking a horizontal average 
($\overline{\bm{V}_{\rm c}}=0, \overline{(\ldots)'}=0$) of equation(\ref{eq:mom0}),
and assuming the Boussinesq approximation for convection motions, we have
\begin{equation}
\rho{d\bm{u}\over d t}+\rho\overline{\bm{V}_{\rm c}\cdot\nabla\bm{V}_{\rm c}}
= -\nabla P - \rho\nabla\psi,
\label{eq:mom_mean2}
\end{equation}
where the second term on the left-hand-side is the turbulent stress.
where we have defined
\begin{equation}
{d\over d t} = {\partial\over\partial t} + \bm{u}\cdot\nabla.
\end{equation}

Similarly, taking horizontal averages of equations (\ref{eq:mass0})
and (\ref{eq:energy0}) we obtain
\begin{equation}
{d\rho\over dt} + \rho\nabla\cdot\bm{u}=0.
\label{eq:mass_mean}
\end{equation}
and
\begin{equation}
\rho T{dS\over dt} + \rho T\overline{\bm{V}_{\rm c}\cdot\nabla S'} 
= \rho\epsilon_{\rm n}  - \nabla\cdot\bm{F}_{\rm rad}, 
\label{eq:energy_mean0}
\end{equation}
where we have assumed that $|T'|\ll T$ and $|\rho' | \ll \rho$.
Since $\nabla\cdot\bm{V}_{\rm c}=0$ (Boussinesq approximation) and hence
$\bm{V}_{\rm c}\cdot\nabla S' = \nabla\cdot(\bm{V}_{\rm c}S')$,
we approximately write equation (\ref{eq:energy_mean0}) as
\begin{equation}
\rho T{dS\over dt} 
= \rho\epsilon_{\rm n}  - \nabla\cdot\bm{F}_{\rm rad} - \nabla\cdot\bm{F}_{\rm con}, 
\label{eq:energy_mean1}
\end{equation}
with defining the convective flux as
\begin{equation}
\bm{F}_{\rm con} \equiv \rho T\overline{\bm{V}_{\rm c}S'} \quad = \rho C_p\overline{\bm{V}_{\rm c} T'}
\label{eq:fconv} 
\end{equation}
\citep{Unno1967}.
Considering the pulsational Lagrangian perturbations $\delta$ and keeping only linear terms we have
\begin{equation}
\delta\bm{F}_{\rm con}
=\bm{F}_{{\rm con} 0}\left({\delta\rho\over\rho} + {\delta T\over T}\right) 
+ \rho T [\overline{(\delta\bm{V}_{\rm c})S'}+ \overline{\bm{V}_{\rm c}\delta S'}].  
\label{eq:dfconv} 
\end{equation}
In the equilibrium condition the mean convective flux has only radial direction,
\begin{equation}  
\bm{F}_{\rm con 0}= {\bf e}_r \rho T \overline{V_{\rm c}^rS'}.
\label{eq:Fc0} 
\end{equation}
We can write equation (\ref{eq:dfconv}) as
\begin{equation}
{\delta\bm{F}_{\rm con}\over F_{\rm con 0}}
=\left({\delta\rho\over\rho} + {\delta T\over T}\right){\bf e}_r 
+ {\delta\bm{V}_{\rm c}\over V_{\rm c}^r}+ 
{\overline{\bm{V}_{\rm c}\delta S'}\over V_{\rm c}^rS'}  
\label{eq:dfconv2} 
\end{equation}
For radial pulsations, $\bm{u}= u\bf{e}_r$, we have
\begin{equation}
{\delta F_{\rm con}\over F_{\rm con}} = \left({\delta\rho\over\rho} + {\delta T\over T}\right)
+ {\delta V_{\rm c}^r\over V_{\rm c}^r}+ {\delta S'\over S'}   
\label{eq:dfconv_r}
\end{equation} 
Then, we have 
\begin{equation}
\delta(\nabla\cdot\bm{F}_{\rm con})
={1\over r^2}{d (r^2 \delta F_{\rm con})\over dr}
-{d\xi_r\over dr}{dF_{\rm con}\over dr}
-{2\xi_r\over r^2}F_{\rm con},
\label{eq:divFc}
\end{equation}
where $\xi_r$ stands for the radial displacement of pulsation.

\subsection{Equations for time-dependent convection}
\subsubsection{Mechanical equations}
Subtracting equation (\ref{eq:mom_mean2}) from equation (\ref{eq:mom0}) and
neglecting a nonlinear term $\rho\bm{u}\cdot\nabla\bm{u}$, we obtaine
\begin{equation}
\rho{d\bm{V}_{\rm c}\over dt} + \rho(\bm{V}_{\rm c}\cdot\nabla\bm{V}_{\rm c}-
\overline{\bm{V}_{\rm c}\cdot\nabla\bm{V}_{\rm c}})
+\rho\bm{V}_{\rm c}\cdot\nabla\bm{u}= -\nabla P' - \rho'\nabla\psi.
\label{eq:mom_conv}
\end{equation}
Here we have disregard $\rho'$ in the left hand side of equation
(\ref{eq:mom0}), because we are using the Boussinesq approximation for convective motion.
Following \citet{Unno1967}'s conjecture that convection motion would approach to equilibrium in
a time scale of the turnover time of convective eddies, we assume that
\begin{equation}
\bm{V}_{\rm c}\cdot\nabla\bm{V}_{\rm c}-\overline{\bm{V}_{\rm c}\cdot\nabla\bm{V}_{\rm c}} = 
\alpha\left({1\over\tau}+\sigma_{\rm r}\right)\bm{V}_{\rm c},
\label{eq:assum_V}
\end{equation}
where $\alpha$ is a numerical parameter of order one, and
$\tau$ is the turnover time of convective eddies, which may be written as
\begin{equation}
\tau = \ell/\varv
\end{equation}
with mixing length $\ell$ and $\varv\equiv\overline{|V_{\rm c}^r|}$. 

We have included $\sigma_{\rm r}$ (the real part of pulsation frequency)
in equation (\ref{eq:assum_V}) in addition to \citet{Unno1967}'s original 
$1/\tau$ term, 
in order to prevent rapid spacial oscillations in pulsation amplitude
that occur near the bottom of the convection zone 
(where $\tau\gg1/\sigma_{\rm r}$) \citep{Gonczi1980,Liege2005}.    
This modification comes from the conjecture that 
if the pulsation period is much shorter
than the turn-over time $\tau$ (associated with largest eddies), 
smaller scale eddies having time-scales of 
$\sim1/\sigma_{\rm r}$ would be important \citep{Saio1980}.
Similar assumption was adopted by \citet{Liege2005}.
Substituting equation\,(\ref{eq:assum_V}) into equation\,(\ref{eq:mom_conv})
we obtain
\begin{equation}
{d\bm{V}_{\rm c}\over dt} = -\alpha\left({1\over\tau}
 +\sigma_{\rm r}\right)\bm{V}_{\rm c} - \bm{V}_{\rm c}\cdot\nabla\bm{u}
 -{1\over\rho}\nabla P' - {\rho'\over\rho}\nabla\psi.
\label{eq:mom_conv1}
\end{equation}
For mass conservation of convection motion, we adopt Boussinesq approximation
\begin{equation}
\nabla\cdot\bm{V}_{\rm c} = 0 .
\label{eq:div0}
\end{equation}

\subsubsection{Thermal equations}
Subtracting equation(\ref{eq:energy_mean0}) from equation(\ref{eq:energy0}) we obtain
\begin{equation}
\begin{array}{l} \displaystyle
\rho T\left[{dS'\over dt} +\bm{V}_{\rm c}\cdot\nabla S\right] 
+ (\rho T)'{dS\over dt} 
\cr \displaystyle \qquad
+\rho T(\bm{V}_{\rm c}\cdot\nabla S'
- \overline{\bm{V}_{\rm c}\cdot\nabla S'})
= (\rho\epsilon_{\rm n})' - \nabla\cdot\bm{F}_{\rm rad}'
\end{array}
\label{eq:energy_conv0}
\end{equation} 
Similarly to equation\,(\ref{eq:assum_V}), we assume that
\begin{equation}
\bm{V}_{\rm c}\cdot\nabla S' 
- \overline{\bm{V}_{\rm c}\cdot\nabla S'} 
= \beta\left({1\over\tau}+\sigma_{\rm r}\right) S',
\end{equation}
where $\beta$ is another numerical parameter of order one.
Then we obtain
\begin{equation}
{dS'\over dt} +\bm{V}_{\rm c}\cdot\nabla S 
 + \left({T'\over T}+{\rho'\over\rho}\right){dS\over dt} 
 +\beta\left({1\over\tau}+\sigma_{\rm r}\right) S'
=- {1\over\rho T}\nabla\cdot\bm{F}'_{\rm rad}.
\label{eq:energy_conv1}
\end{equation} 
where we neglected the term $(\rho\epsilon_{\rm n})'$ because
we are interested only in the envelope convection.
For the last term of the above equation we assume
\begin{equation}
\nabla\cdot\bm{F}'_{\rm rad}\approx -{4acT^3\over 3\kappa\rho}\nabla^2 T'
\approx {4acT^3\over 3\kappa\rho}{T'\over (\ell^2/f)},
\end{equation}
where $\kappa$ is opacity per unit mass, $\ell$  mixing length, 
and $f$ is a numerical factor; its appropriate value
will be discussed in the next subsection.

Furthermore, according to the assumptions in the mixing-length theory, 
we assume $P' \approx 0$ in obtaining convective perturbations of 
the thermodynamic relations 
(but $P'$ is retained in the equation of motion discussed below). 
Then, we have
\begin{equation}
{T'\over T} \approx \left(\partial\ln T\over\partial S\right)_P S' = {S'\over C_p},
\end{equation}
and
\begin{equation}
\nabla\cdot\bm{F}'_{\rm rad} 
\approx {4acT^4\over 3\kappa\rho C_p}{S'\over (\ell^2/f)}
=\rho TK_{\rm rad}S', 
\end{equation} 
with
\begin{equation}
K_{\rm rad}\equiv {4acT^3\over 3\kappa\rho^2 C_p (\ell^2/f)}.
\label{eq:Krad}
\end{equation}
Under the same assumption, we can write the density perturbation as 
\begin{equation}
{\rho'\over\rho} \approx 
\left(\partial\ln\rho\over\partial\ln T\right)_P{T'\over T} =
-{\chi_T\over\chi_\rho}{T'\over T} = -{\chi_T\over\chi_\rho}{S'\over C_p}.
\label{eq:rhoS}
\end{equation}
Then we obtain
\begin{equation}
\rho T' + T\rho' \approx \rho T\left(1-{\chi_T\over\chi_\rho}\right)
{S'\over C_p}, 
\end{equation}
where $\chi_T\equiv(\partial\ln P/\partial\ln T)_\rho$ and
$\chi_\rho\equiv(\partial\ln P/\partial\ln\rho)_T$.
Using these relations in eq.\,(\ref{eq:energy_conv1}), we obtain
\begin{equation}
{dS'\over dt} +\bm{V}_{\rm c}\cdot\nabla S 
+ \left(1-{\chi_T\over\chi_\rho}\right){S'\over C_p}{dS\over dt} 
+\left({\beta\over\tau}+\beta\sigma_{\rm r}+K_{\rm rad}\right)S'
=0.
\label{eq:energy_conv2}
\end{equation}

\subsubsection{Convection in equilibrium}
In this subsection we discuss the relation between our formulations and the mixing-length theory to obtain relations among the parameters $\alpha$, $\beta$ and 
convective eddy shapes.
If there is no pulsation we have steady-state convection, in which $\bm{V}_{\rm c0}$ is
given from equation (\ref{eq:mom_conv1})
\begin{equation}
{\alpha\over\tau}\bm{V}_{{\rm c}0}=-{1\over\rho}\nabla P'_0 - {\rho'_0\over\rho}
\nabla\psi. 
\end{equation}
Treating the convective motion locally by replacing $\nabla$ with $i\bm{k}$, 
we obtain the radial component of convection velocity $V_{\rm c0}^r$
\begin{equation}
{\alpha\over\tau}V_{{\rm c}0}^r=-ik_r\left(P'\over\rho\right)_0 
- g\left({\rho'\over\rho}\right)_0 
\label{eq:stconv_r}
\end{equation}
and the horizontal component $\bm{V}_{\rm c0}^h$
\begin{equation} 
\bm{V}_{{\rm c}0}^h = -i{\tau\over\alpha}\bm{k}_h\left(P'\over\rho\right)_0.
\label{eq:stconv_h}
\end{equation}
The continuity equation $\nabla\cdot\bm{V}_{{\rm c}0} = 0$ can be written as
\begin{equation}
k_rV_{{\rm c}0}^r+\bm{k}_h\cdot\bm{V}_{{\rm c}0}^h = 0.
\label{eq:eqconv}
\end{equation}
Combining equations (\ref{eq:stconv_h}) and (\ref{eq:eqconv}), we obtain
\begin{equation}
ik_r\left(P'\over\rho\right)_0={k_r^2\over k_h^2}{\alpha V_0^r\over \tau} .
\end{equation}
Substituting this equation into the first relation in equation\,(\ref{eq:stconv_r}), 
we obtain
\begin{equation}
\left({\rho'\over\rho}\right)_0=-{\alpha\over\tau g}{k^2\over k_h^2}V_{{\rm c}0}^r
\quad {\rm or} \quad
\left({S'\over C_p}\right)_0{\chi_T\over\chi\rho}={\alpha\over\tau g}{k^2\over k_h^2}V_{{\rm c}0}^r,
\label{eq:st_rho_s}
\end{equation}
where $k^2 = k_r^2 + k_h^2$.

In the equilibrium state eq.\,(\ref{eq:energy_conv2}) can be written as
\begin{equation}
V_{{\rm c}0}^r{dS_0\over dr} 
= -\left({\beta\over\tau}+K_{\rm rad}\right)S'_0.
\label{eq:energy_convequ}
\end{equation} 
Since 
\begin{equation}
{dS_0\over dr} = - {C_p\over H_p}(\nabla - \nabla_{\rm ad}),
\end{equation}
equation (\ref{eq:energy_convequ}) may be written as
\begin{equation}
 \left({\beta\over\tau} + K_{\rm rad}\right)S'_0
= V_{{\rm c}0}^r{C_p\over H_p}(\nabla-\nabla_{\rm ad}). 
\label{eq:S-Vr}
\end{equation}
A relation with the mixing length theory is apparent if we have the 
horizontal average of the absolute value and use the relation 
$\overline{|V_{{\rm c}0}^r|} = \varv = \ell/\tau$,
\begin{equation}
{\overline{|S'_0|}\over C_p}={\overline{|T'_0|}\over T} = {1\over\beta}{1\over 1 +(\tau K_{\rm rad}/\beta)}
(\nabla-\nabla_{\rm ad}){\ell\over H_p}.
\label{eq:deltaT}
\end{equation} 
In the formulation of MLT by \citet{Henyey1965} 
the efficiency factor $\gamma$ defined as
\begin{equation}
\gamma = {\nabla-\nabla'\over\nabla'-\nabla_{\rm ad}} 
\qquad \rightarrow \qquad
\nabla - \nabla' = {\gamma\over\gamma+1}(\nabla-\nabla_{\rm ad}),
\label{eq:gamma}
\end{equation}
where $\nabla'$ is the temperature gradient felt by moving eddies (sometimes called 
internal temperature gradient).

According to eq.(36) in \citet{Henyey1965}(their opacity per unit volume is converted to the opacity per unit mass here),
 for optically thick convective eddies,
\begin{equation}
{1\over \gamma} = {2acT^3\over C_p\varv\kappa\rho^2\ell y}={3\over 2}K_{\rm rad}{\tau\over y f}
\end{equation}
with $y=3/(4\pi^2)$. If we set $f=2\pi^2\beta$, we have 
$\tau K_{\rm rad}/\beta = 1/\gamma$. Then equation (\ref{eq:deltaT}) becomes 
\begin{equation}
{\overline{|T_0'|}\over T} = {1\over\beta}{\gamma\over \gamma +1}
(\nabla-\nabla_{\rm ad}){\ell\over H_p}={1\over\beta}(\nabla-\nabla'){\ell\over H_p}.
\label{eq:deltaT2}
\end{equation}
This equation can be derived from MLT, where $\beta =2$ is usually adopted.
Using equation (\ref{eq:deltaT2}) in equation\,(\ref{eq:st_rho_s}) we have
\begin{equation}
\left(V_{{\rm c}0}^r\right)^2 = {1\over\beta\alpha}\left(1-{2\over 3}Q\right)gH_p
{\chi_T\over\chi_\rho}(\nabla-\nabla')\left({\ell\over H_p}\right)^2,
\label{eq:v02rv}
\end{equation}
where $Q$ is defined as $k_r^2/k^2 = 2Q/3$ meaning $Q=1$ for isotropic eddies
\citep{Gabriel1975}.
In MLT, 
\begin{equation}
{1\over\beta\alpha}\left(1-{2\over 3}Q\right) = 1/8 
\end{equation}
is often adopted \citep[e.g.][]{Henyey1965}, which corresponds to the relation
\begin{equation}
\alpha = {8\over\beta}\left(1-{2\over 3}Q\right).
\end{equation}
Therefore, we can adopt $(Q,\alpha,\beta)=(1, 1.3, 2)$ as a standard set of parameters.

\subsection{Pulsational perturbation on convection}
In this subsection we consider spherical symmetric (radial) linear Lagrangian 
perturbation $\delta$ for convective eddies which, as in the previous subsection,
are treated locally. We will present 
$\delta V^r$ and $\delta S'$ (convection variables perturbed by pulsation)
as functions of regular pulsation variables.

\subsubsection{Thermal relations}
Applying pulsational perturbations to equation\,(\ref{eq:energy_conv2}) 
and using equation\,(\ref{eq:energy_convequ}),
we obtain
\begin{equation}
\begin{array}{ll}\displaystyle
\Sigma_1{\delta S'\over S'} - K_{\rm rad}{\delta V^r\over V^r}
=
\left({\beta\over\tau}+K_{\rm rad}\right){\delta(dS/dr)\over(dS/dr)}
\cr \displaystyle \qquad \qquad
-i\sigma\left(1-{\chi_T\over\chi_\rho}\right){\delta S\over C_p} 
-\delta K_{\rm rad}+{\beta\over\tau}\left({\delta H_p\over H_p}\right),
\end{array}
\label{eq:energy_convpul2}
\end{equation} 
where we defined
\begin{equation}
\Sigma_1\equiv\left(i\sigma+{\beta\over\tau}+
\beta\sigma_{\rm r}+K_{\rm rad}\right)
\label{eq:Sigma1}
\end{equation}

\subsubsection{Mechanical relations}
From the momentum equation\,(\ref{eq:mom_conv1}) of convective eddies, we obtain
\begin{equation}
\begin{array}{l} \displaystyle
\left(i\sigma+{\alpha\over\tau}+\alpha\sigma_{\rm r}\right)\delta\bm{V}_{\rm c}  
= {\alpha\over\tau}{\delta\tau\over\tau}\bm{V}_{{\rm c}0}- i\sigma(\bm{V}_{{\rm c}0}\cdot\nabla)\bm{\xi}
 \cr \displaystyle \qquad \qquad \qquad
 -i\bm{k}\delta\left(P'\over\rho\right) 
 - \delta\left({\rho'\over\rho}\right)\nabla\psi
 -{\rho'\over\rho}\delta(\nabla\psi).
\end{array}
\label{eq:mom_pul}
\end{equation}
From equation\,(\ref{eq:div0}) we obtain
\begin{equation}
\bm{k}\cdot\delta\bm{V}_{\rm c}=0.
\label{eq:div0_pul}
\end{equation} 
Taking the inner product between $\bm{k}$ and equation\,(\ref{eq:mom_pul})
and using equation\,(\ref{eq:div0_pul}) we have
\begin{equation}
0=- i\sigma\bm{k}\cdot\left[(\bm{V}_{{\rm c}0}\cdot\nabla)\bm{\xi}\right]
 -ik^2\delta\left(P'\over\rho\right) 
 - k_r\delta\left({\rho'\over\rho}\right){d\psi\over dr}
 -{\rho'\over\rho}k_r\delta\left({d\psi\over dr}\right),
\end{equation}
Substituting above equation into eq.\,(\ref{eq:mom_pul}), we eliminate $P'$,
\begin{equation}
\begin{array}{ll}\displaystyle
\left(i\sigma+{\alpha\over\tau}+\alpha\sigma_{\rm r}\right)\delta\bm{V}_{\rm c} 
=i\sigma\left[{\bm{k}\over k^2}\bm{k}\cdot(\bm{V}_{\rm c0}\cdot\nabla)\bm{\xi}
 -(\bm{V}_{\rm c0}\cdot\nabla)\bm{\xi}\right] 
 \cr\displaystyle  \qquad
-g{\chi_T\over\chi_\rho}
\left({k_r\bm{k}\over k^2}-{\bf e}_r\right){\delta S'\over C_p}
+{\alpha\over\tau}\left({\delta H_p\over H_p}
-{\delta\varv\over\varv}\right)\bm{V}_{\rm c0} 
  \cr\displaystyle  \qquad
-{\alpha V_{\rm c0}^r\over\tau g}{k^2\over k_h^2} 
\left[{\bm{k}\over k^2}\bm{k}\cdot\delta(\nabla\psi)
 -\delta(\nabla\psi)\right],
\end{array}
\label{eq:deltV}
\end{equation}
where we used the relations 
$\tau = \ell/\varv$ \rm and $\ell \propto H_p$, 
and the relation between $\rho'$ and $S'$ (eq.\ref{eq:rhoS}).

The radial component of equation\,(\ref{eq:deltV}) is can be written as
\begin{equation}
\begin{array}{ll}\displaystyle
\Sigma_2{\delta V^r_{\rm c}\over V_{\rm c}^r} 
-{\alpha\over\tau}{\delta S'\over S'}
= i\sigma\left[{2Q\over3}\left({d\xi\over dr}-{\xi\over r}\right)
 -{d\xi\over dr}\right] 
\cr \displaystyle \qquad \qquad\qquad
+{\alpha\over\tau}{\delta H_p\over H_p}
+{\alpha\over\tau g}\delta\left({d\psi\over dr}\right),
\end{array}
\label{eq:deltVr_v2}
\end{equation}
where we defined
\begin{equation}
\Sigma_2\equiv i\sigma+{2\alpha\over\tau}+\alpha\sigma_{\rm r}.
\label{eq:Sigma2}
\end{equation}
In deriving equation\,(\ref{eq:deltVr_v2}),
we assumed the relation $\delta\varv/\varv = \delta V_{\rm c}^r/V_{\rm c0}^r$, 
and used the relation between $S_0'$ and $V_{{\rm c}0}^r$ given in equation\,(\ref{eq:st_rho_s}). 
We also used the relation
\begin{equation}
{k_r\over k^2}\bm{k}\cdot(\bm{V}_{\rm c0}\cdot\nabla)\bm{\xi}
={2Q\over3}V_{\rm c0}^r\left({d\xi\over dr}-{\xi\over r}\right),
\end{equation}
which can be derived from the relations $\bm{k}\cdot\bm{V}_{\rm c0} = 0$
(eq.\,\ref{eq:eqconv})
and $k_r^2/k^2= 2Q/3$ (eq.\,\ref{eq:v02rv}).

From equations\,(\ref{eq:energy_convpul2}) and (\ref{eq:deltVr_v2})
we can express $\delta V^r_{\rm c}/V^r_{\rm c}$ and $\delta S'/S'$ 
as functions of pulsation variables, which are used 
in equation\,(\ref{eq:Lcon}) (below) to obtain the Lagrangian 
perturbation of convective luminosity, $\delta L_{\rm conv}$.


\subsection{Equations for linear radial pulsations}
Using radial displacement $\xi$, the Lagrangian perturbations of pressure 
$\delta P$, and entropy $\delta S$, 
linearised momentum (\ref{eq:mom_mean2}) and 
mass conservation (\ref{eq:mass_mean}) equations can be written as
\begin{equation}
r{d\over dr}\left({\xi_r\over r}\right)= -3{\xi_r\over r} - {\delta\rho\over\rho} 
= -3{\xi_r\over r} - {1\over\Gamma_1}{\delta p\over p} +{\chi_T\over\chi_\rho}{\delta S\over C_p}
\label{eq:dxidr}
\end{equation}
and
\begin{equation}
{d\over d\ln r}\left(\delta p\over p\right) =
V\left({\sigma^2r\over g}+4\right){\xi_r\over r}+ V{\delta p\over p}, 
\label{eq:delpdr}
\end{equation}
in which we have neglected 
turbulent pressure term $(\overline{\bm{V}_{\rm c}\cdot\nabla\bm{V}_{\rm c}})'_r$.

\subsubsection{Energy conservation}
Linearized energy conservation may be written as
\begin{equation}
i\sigma T\delta S 
= \epsilon_{\rm n}\left(\epsilon_\rho{\delta\rho\over\rho}
+\epsilon_T{\delta T\over T}\right)
-L_r{d\over dM_r}\left(\delta L_r\over L_r\right) 
- \left(\delta L_r\over L_r\right){dL_r\over dM_r},
\end{equation}
where $\epsilon_\rho\equiv(\partial\ln\epsilon_{\rm n}/\partial\ln\rho)_T$ and
$\epsilon_T\equiv(\partial\ln\epsilon_{\rm n}/\partial\ln T)_\rho$.
Luminosity perturbation can be expressed as
\begin{equation}
{\delta L_r\over L_r} = {L_{\rm rad}\over L_r}{\delta L_{\rm rad}\over L_{\rm rad}} + {L_{\rm con}\over L_r}{\delta L_{\rm con}\over L_{\rm conv}}.
\end{equation}
Radiative luminosity $L_{\rm rad}$ and it's Lagrangian perturbation 
$\delta L_{\rm rad}$ are given as
\begin{equation}
L_{\rm rad} = -(4\pi r^2)^2{4acT^4\over 3\kappa}{d\ln T\over dM_r},
\end{equation}
and
\begin{equation}
{\delta L_{\rm rad}\over L_{\rm rad}}
=4{\xi_r\over r}+4{\delta T\over T} -{\delta\kappa\over\kappa} 
+ {1\over d\ln T/dM_r}{d\over dM_r}\left(\delta T\over T\right).
\end{equation}

For the perturbation of convective luminosity we have 
from equation\,(\ref{eq:dfconv_r})
\begin{equation}
{\delta L_{{\rm con}}\over L_{\rm con}} = 2{\xi_r\over r} + {\delta F_{{\rm con}}\over F_{\rm con}}
= 2{\xi_r\over r} + {\delta\rho\over\rho} + {\delta T\over T}
+ {\delta V^r\over V^r}+ {\delta S'\over S'}.
\label{eq:Lcon}
\end{equation}
The convection variables 
${\delta V^r/V^r}$ and ${\delta S'/S'}$
can be expressed by ordinary pulsation variables using
equations\,(\ref{eq:energy_convpul2}) and (\ref{eq:deltVr_v2}).

\subsection{Summary of the linear differential equations for radial pulsations}
We define non-dimensional variables $Z_1 \ldots Z_4$ as
\begin{equation}
Z_1={\delta P\over P}, \quad Z_2={\delta S\over C_p}, \quad Z_3 = {\delta r\over r},
\quad Z_4 = {\delta L_r\over L_r}.
\end{equation}
The differential equations may be given as
\begin{equation}
\begin{array}{ll}\displaystyle
{dZ_1\over d\ln r} = VZ_1 + V\left(c_1\omega^2+4\right)Z_3,
\cr\cr \displaystyle
{dZ_2\over d\ln r} = b_1Z_1 + b_2Z_2 + b_3Z_3 -b_4{L_r\over L_{\rm rad}}Z_4,
\cr\cr \displaystyle
{dZ_3\over d\ln r}=  - {1\over\Gamma_1}Z_1 +{\chi_T\over\chi_\rho}Z_2 - 3Z_3,
\cr\cr \displaystyle
{dZ_4\over d\ln r}= b_5Z_1 + b_6Z_2 - {d\ln L_r\over d\ln r}Z_4,
\end{array}
\end{equation}
where $V\equiv -d\ln\,P/d\ln\,r$, $c_1\equiv(r/R)^3M/M_r$, 
and the pulsation angular frequency $\sigma$ is normalized as 
$\omega=\sigma\sqrt{R^3\over GM}$. 
Coefficients $b_1 \ldots b_6$ are defined as
\begin{equation*}
\begin{array}{ll}\displaystyle
b_1 = b_4\left[(4-\kappa_T)\nabla_{\rm ad}-{\kappa_\rho\over\Gamma_1}
+{\nabla_{\rm ad}\over \nabla}\left({d\ln\nabla_{\rm ad}\over d\ln P}-1\right)
\right. \cr \displaystyle \qquad \qquad \qquad \left. 
+f_{\rm CR}\left(\nabla_{\rm ad}+{1\over \Gamma_1}+a_p\right)\right],
\cr\cr \displaystyle
b_2 = b_4\left[ 4-\kappa_T +\kappa_\rho{\chi_T\over\chi_\rho}+ f_{\rm CR}\left(1-{\chi_T\over\chi_\rho} 
+a_s\right)\right],
\cr\cr \displaystyle
b_3 = b_4\left[4-{\nabla_{\rm ad}\over\nabla}\left(c_1\omega^2+4\right)+2f_{\rm CR}\left(1+a_r\right)\right],
\cr\cr \displaystyle
b_4 =\left[{1\over V\nabla}+{f_{\rm CR}\Sigma_{23}\over\Sigma_4}
{K_{\rm rad}+\beta/\tau
\over V(\nabla-\nabla_{\rm ad})}\right]^{-1},
\cr\cr \displaystyle
b_5 = {4\pi r^3\rho\epsilon\over L_r}\left(\epsilon_T\nabla_{\rm ad}+{\epsilon_\rho\over\Gamma_1}\right),
\cr\cr \displaystyle
b_6 = {4\pi r^3\rho\epsilon\over L_r}\left(\epsilon_T-\epsilon_\rho{\chi_T\over\chi_\rho}\right) - i\omega\sqrt{GM\over R^3}{4\pi r^3 TC_p\over L_r},
\end{array}
\end{equation*}
where 
$f_{\rm CR}\equiv L_{\rm conv}(r)/L_{\rm rad}(r),$
\begin{equation*}
\begin{array}{ll}\displaystyle
a_r ={1\over \Sigma_4}\left[i\sigma\Sigma_5v_{rr}
+\Sigma_{23}\left(3K_{\rm rad}+{\beta\over\tau}\right)
+{1\over\tau}\left(\alpha\Sigma_5+\beta\Sigma_{23}\right)\right],
\cr\cr\displaystyle
a_p={1\over \Sigma_4}\left[
i\sigma\Sigma_5{v_{r\rho}\over\Gamma_1}
+\Sigma_{23}s_p+{1\over\tau}(\alpha\Sigma_5+\beta\Sigma_{23})\left(1-{1\over\Gamma_1}\right)
\right],
\cr\cr\displaystyle
a_s={1\over \Sigma_4}\left[-i\sigma\Sigma_5v_{r\rho}{\chi_T\over\chi_\rho}
+\Sigma_{23}s_s
+{1\over\tau}(\alpha\Sigma_5+\beta\Sigma_{23}){\chi_T\over\chi_\rho}\right],
\end{array}
\end{equation*}
\begin{equation*}
\begin{array}{ll}\displaystyle
\Sigma_{23} \equiv \Sigma_2 +{\alpha\over\tau}, \quad
\Sigma_4 \equiv \Sigma_1\Sigma_2-K_{\rm rad}{\alpha\over\tau},
\quad \Sigma_5 \equiv \Sigma_1+K_{\rm rad},
\cr\cr \displaystyle
v_{rr} \equiv 1-Q-{\alpha\over i\sigma\tau}, 
\qquad
v_{r\rho} \equiv  1-{2Q\over3}, 
\end{array}
\end{equation*}
\begin{equation*}
\begin{array}{ll}\displaystyle
s_p \equiv {1\over\Gamma_1}\left(K_{\rm rad}+{\beta\over\tau}\right)
-K_{\rm rad}\left[(3-\kappa_T)\nabla_{\rm ad}-2-{\kappa_\rho \over \Gamma_1}\right],
\cr\cr\displaystyle
s_s \equiv -\left(K_{\rm rad}+{\beta\over\tau}\right)
\left[{\chi_T\over\chi_\rho}-{d\ln C_p/d\ln p\over \nabla-\nabla_{\rm ad}}\right] 
\cr \qquad \qquad \displaystyle
+\left[i\sigma\left({\chi_T\over\chi_\rho}-1\right)-K_{\rm rad}\left(3-\kappa_T +\kappa_\rho{\chi_T\over\chi_\rho}\right)\right].
\end{array}
\end{equation*}


\bsp	
\label{lastpage}
\end{document}